\documentclass[10pt,twoside,twocolumn,english,superscriptaddress, aps]{revtex4-1}
\usepackage[T1]{fontenc}
\usepackage[latin9]{inputenc}
\usepackage{color}
\usepackage{babel}
\usepackage{verbatim}
\usepackage{textcomp}
\usepackage{amsmath}
\usepackage{graphicx}
\usepackage[unicode=true]
 {hyperref}

\makeatletter

\newcommand{\lyxmathsym}[1]{\ifmmode\begingroup\def\b@ld{bold}
  \text{\ifx\math@version\b@ld\bfseries\fi#1}\endgroup\else#1\fi}

\providecommand{\tabularnewline}{\\}
\newcommand{\lyxdot}{.}

\@ifundefined{date}{}{\date{}}
\usepackage{makecell}
\date{\today}

\@ifundefined{showcaptionsetup}{}{%
 \PassOptionsToPackage{caption=false}{subfig}}
\usepackage{subfig}
\makeatother

\begin{document}
\author{Y. Lu}
\affiliation{Department of Physics and Astronomy, Rice University, Houston, Texas 77005, USA}
\affiliation{Theoretical Division, Los Alamos National Laboratory, Los Alamos, New Mexico, 87545, USA}
\author{P. Tzeferacos}
\affiliation{Department of Astronomy and Astrophysics, University of Chicago, Chicago, Illinois 60637, USA}
\author{E. Liang}
\affiliation{Department of Physics and Astronomy, Rice University, Houston, Texas 77005, USA}
\author{R. K. Follett}
\affiliation{Laboratory for Laser Energetics, University of Rochester, Rochester, New York 14623, USA}
\author{L. Gao}
\affiliation{Princeton Plasma Physics Laboratory, Princeton, New Jersey 08540, USA}
\author{A. Birkel}
\affiliation{Plasma Science and Fusion Center, Massachusetts Institute of Technology, Cambridge, Massachusetts 02139, USA}
\author{D. H. Froula}
\affiliation{Laboratory for Laser Energetics, University of Rochester, Rochester, New York 14623, USA}
\author{W. Fu}
\affiliation{Department of Physics and Astronomy, Rice University, Houston, Texas 77005, USA}
\author{H. Ji}
\affiliation{Princeton Plasma Physics Laboratory, Princeton, New Jersey 08540, USA}
\author{D. Lamb}
\affiliation{Department of Astronomy and Astrophysics, University of Chicago, Chicago, Illinois 60637, USA}
\author{C.K. Li}
\author{H. Sio}
\author{R. Petrasso}
\affiliation{Plasma Science and Fusion Center, Massachusetts Institute of Technology, Cambridge, Massachusetts 02139, USA}
\author{M. Wei}
\affiliation{Laboratory for Laser Energetics, University of Rochester, Rochester, New York 14623, USA}

\title{Numerical Simulation of magnetized jet creation using a hollow ring
of laser beams}
\begin{abstract}
Three dimensional FLASH magneto-hydrodynamics(MHD) modeling is carried
out to interpret the OMEGA laser experiments of strongly magnetized,
highly collimated jets driven by a ring of 20 OMEGA beams. The predicted
optical Thomson scattering spectra and proton images are in good agreement
with a subset of the experimental data. Magnetic fields generated
via the Biermann battery term are amplified at the boundary between
the core and the surrounding of the jet. The simulation predicts multiple
axially aligned magnetic flux ropes with alternating poloidal component.
Future applications of the hollow ring configuration in laboratory
astrophysics are discussed.

\end{abstract}

\maketitle

\section{Introduction}

Supersonic, well collimated outflows are ubiquitous in many astrophysical
systems, such as young stellar objects (YSO)\citep{YSO_Hirose1997},
active galactic nucleus (AGN)\citep{AGN_Ferrari1998} and gamma-ray
bursts (GRB)\citep{GRB_Sari1999}. Despite various astronomical observations,
theoretical studies, and numerical modelings of astrophysical jets,
many fundamental questions remain, e.g. launching mechanism, composition,
morphology of the magnetic field and stability. Magnetic fields permeate
the universe, but their origin is not fully understood, especially
in astrophysical jets. A variety of ideas have been proposed in which
seed magnetic fields could be created. However, this mechanism has
only been demonstrated in the laboratory recently\citep{Tzeferacos2017,Tzeferacos2018}.

With advances in large laser facilities, scalable laboratory experiments
to study astrophysical phenomena have become achievable. Over the
years, experiments have been performed to study astrophysics utilizing
high-intensity lasers at the OMEGA laser facility at the Laboratory
for Laser Energetics(LLE), and the National Ignition Facility(NIF)
at Lawrence Livermore National Laboratory as well as laser facilities
in other countries\citep{HED_Remington2005,HED_Foster2005,Jet_Ciardi2007,Jet_Ciardi2009,Jet_Lebedev2004,Jet_Lebedev2005,Jets_Lebedev2002,Remington2006}.
Laboratory produced jets with proper dimensionless parameters may
provide an alternative platform to study the jets of astrophysical
scales.

A new way of launching high density and high temperature plasma jets
using multiple intense laser beams is to utilize the hollow ring configuration
as proposed by Fu et. al\citep{Fu_Fu2013,Fu_Fu2015}. It was demonstrated
in two dimensional numerical simulations that a bundle of laser beams
of given individual intensity, duration and focal spot size, produces
a supersonic jet of higher density, temperature and better collimation,
if the beams are focused to form a circular ring pattern on a flat
target instead of a single focal spot. The Biermann Battery ( $\nabla n_{e}\times\nabla T_{e}$
) term\citep{Biermann1950} can generate and sustain strong toroidal
fields downstream in the collimated jet outflow far from the target
surface. Those simulations were carried out in two dimensional cylindrical
geometry, where the intensity variation along the ring due to individual
laser beams were neglected. Three dimensional simulations are necessary
to understand the formation and evolution of the jet in the actual
experiments.

The ring jet experiments were designed and carried out on OMEGA laser
facility\citep{OMEGA_Boehly1997} in 2015 and 2016\footnote{L. Gao et. al. to be submitted}.
We used 20 OMEGA beams to simultaneously irradiate the target forming
a ring pattern. Each beam delivers 500J energy in $1\mathrm{ns}$.
In this paper, we aim to use three dimensional FLASH\citep{FLASH_Fryxell2000}
simulations to explain the observed jet parameters in the experiments.
Using the MHD results, we can predict the diagnostics outcomes from
first principles. In Sec. \ref{sec:Simulation-method}, we describe
the experiment design to produce laboratory jets on the OMEGA laser
and the simulation methods to model the experiment. Simulation results
are discussed in Sec. \ref{sec:FLASH-Simulation-results}. The validation
against a subset of experimental data is discussed in Sec. \ref{sec:Comparison-to-experiments}.

\section{\label{sec:Simulation-method}Simulation methods}

\subsection{Non-ideal magneto-hydrodynamics in FLASH code\label{subsec:Non-ideal-magneto-hydrodynamics-}}

The FLASH code\footnote{FLASH4 is available at \href{https://flash.uchicago.edu/}{https://flash.uchicago.edu/}}
is used to carry out the detailed physics simulations of our laser
experiments to study the formation and dynamics of the jet and the
origin of magnetic fields. FLASH\citep{FLASH_Dubey2009,FLASH_Fryxell2000}
is a publicly available, multi-physics, highly scalable parallel,
finite-volume Eulerian code and framework whose capabilities include:
adaptive mesh refinement (AMR), multiple hydrodynamic and MHD solvers,
implicit solvers for diffusion using the HYPRE library and laser energy
deposition. FLASH is capable of using multi-temperature equation of
states and multi-group opacities. Magnetic field generation via the
Biermann battery term has been implemented and studied in FLASH recently\citep{FLASH_BB_Graziani2015}. 

\begin{table}
\caption{Target characteristics used in the experiments and the simulations\label{tab:Target-characteristics}}

\begin{tabular}{|c|c|c|}
\hline 
\makecell{Composition\\(number fraction)} & Density & Laser target Ring radius\tabularnewline
\hline 
\hline 
C(50\%) H(50\%) & 1.04g/cc & 0, 400, 800, 1200$\mathrm{\mu m}$\tabularnewline
\hline 
C(49\%) H(49\%) Fe(2\%) & 1.21g/cc & 800, 1200$\mathrm{\mu m}$\tabularnewline
\hline 
\end{tabular}
\end{table}

We use the same FLASH code units as in \citep{Tzeferacos2017} to
solve the three-temperature non-ideal MHD equations. A cartesian grid
with $(256\times256\times512)$ zones is used to resolve a $(3\mathrm{mm}\times3\mathrm{mm}\times6\mathrm{mm})$
domain, corresponding to $\sim$11$\mathrm{\mu m}$ per cell width.
The number of cells we use is sufficient to resolve the spatial distribution
of all the quantities that the plasma diagnostics are able to resolve
in the OMEGA experiments. We did test runs at diffrent resolutions,
and the simulation converge at a cell with of 11$\mathrm{\mu m}$.
The plasma has zero initial magnetic field. The laser target is modeled
as a 3mm diameter and 0.5mm thick disk with the composition listed
in Table \ref{tab:Target-characteristics}. To model the material
properties of the CH and CH$+$dopant targets, we utilize the opacity
and EoS tables computed with PROPACEOS\footnote{PROPACEOS is available at \href{http://www.prism-cs.com}{http://www.prism-cs.com}}.
We use the equation of state of helium in the chamber with initial
density equal to $2\times10^{-7}\mathrm{g/cc}$, which should have
been vacuum. The helium does not affect the simulation significantly,
as the mass, momentum and energy budget in the modeled helium is much
less than 1\%. To suppress the magnetic field from numerical artefact,
we turn off the Biermann battery term and use the largest allowed
magnetic resistivity in the explicit solver for each time step in
the regions with density lower than $2\times10^{-5}\mathrm{g/cm^{3}}$.
The electron heat conduction is calculated using Braginskii model\citep{cond_Braginskii1965}
in weak magnetic field limit.

\begin{figure}
\subfloat[]{\includegraphics[scale=0.2]{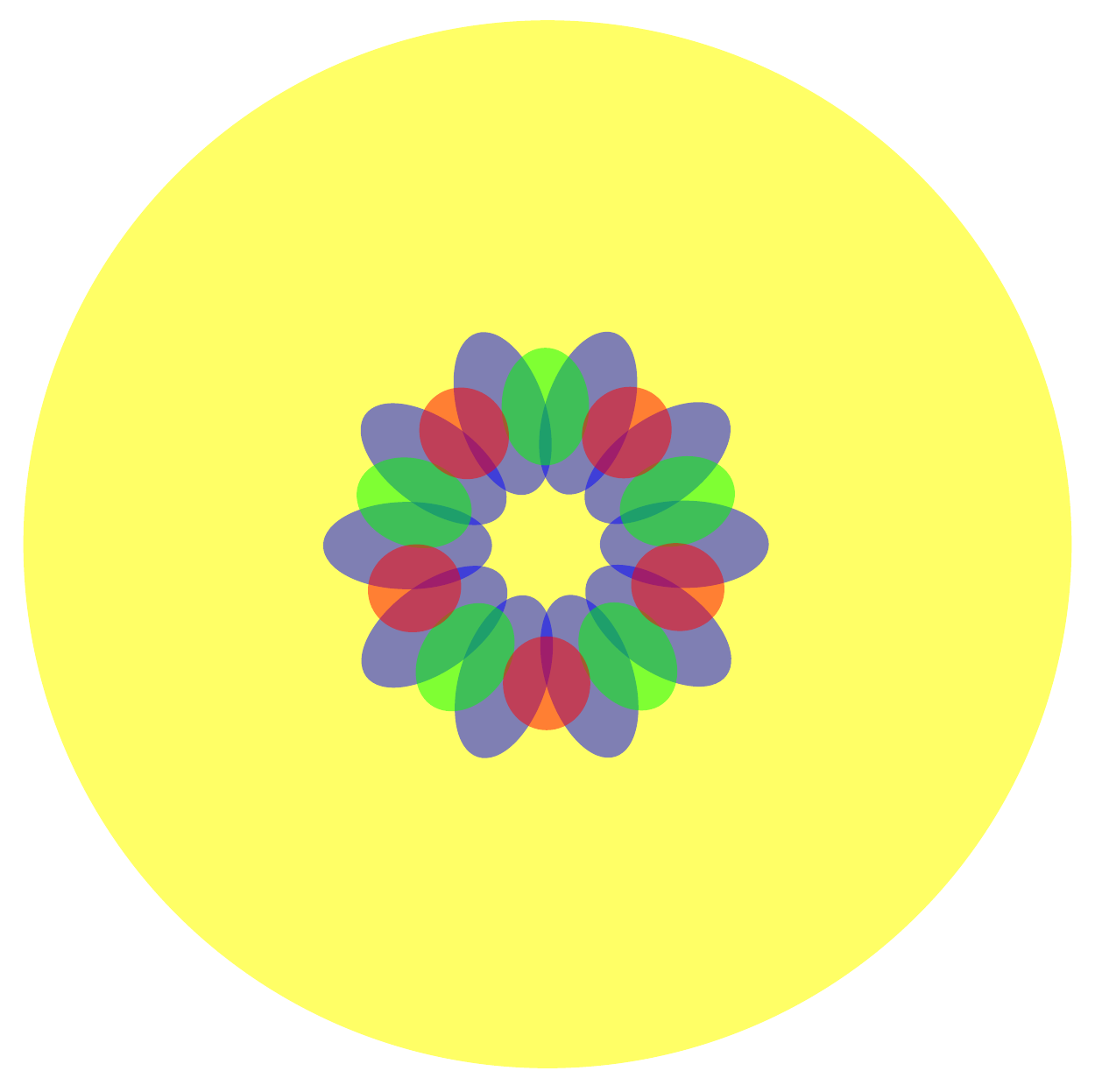}

}\subfloat[]{\includegraphics[scale=0.2]{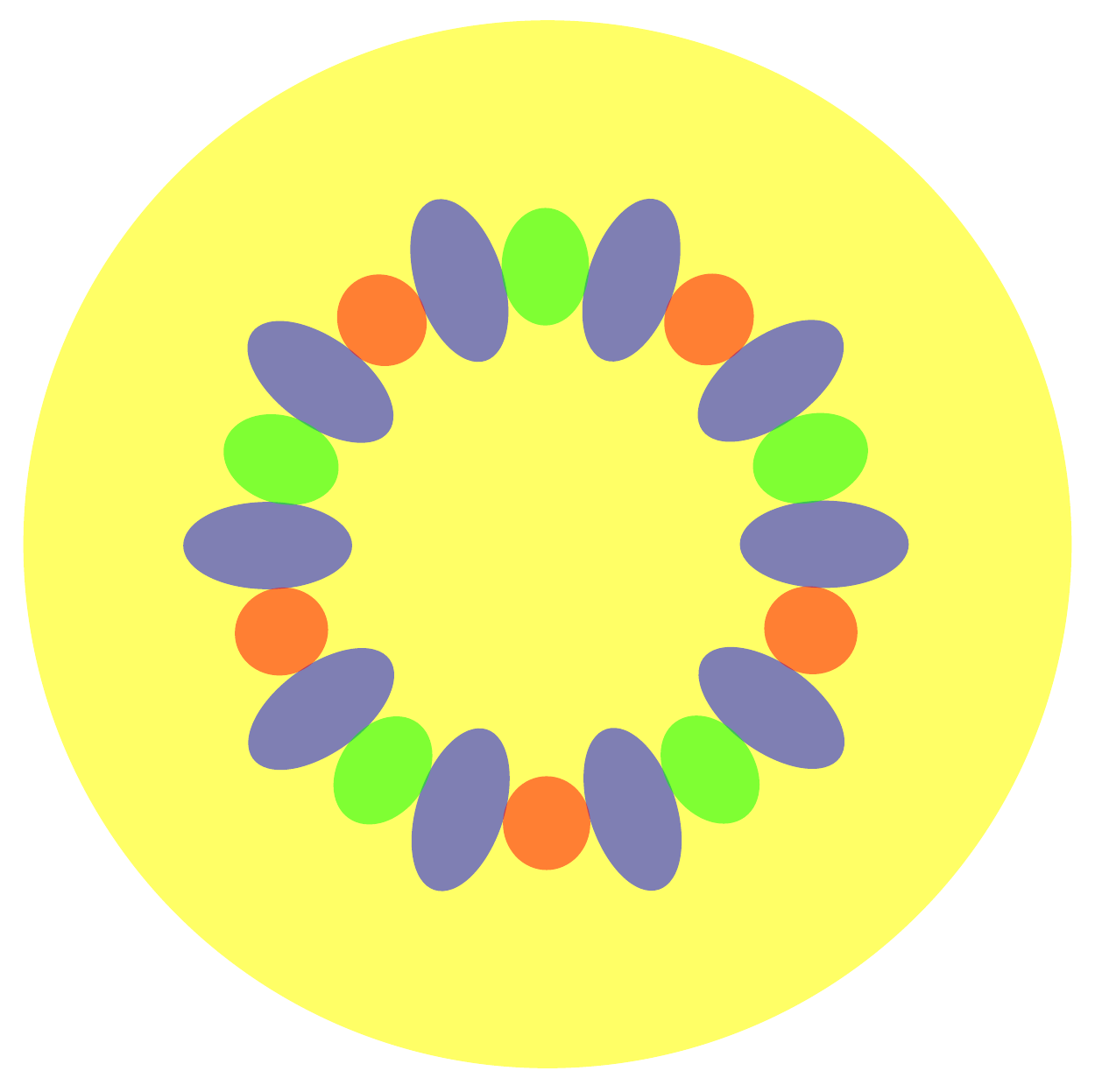}

}\subfloat[]{\includegraphics[scale=0.2]{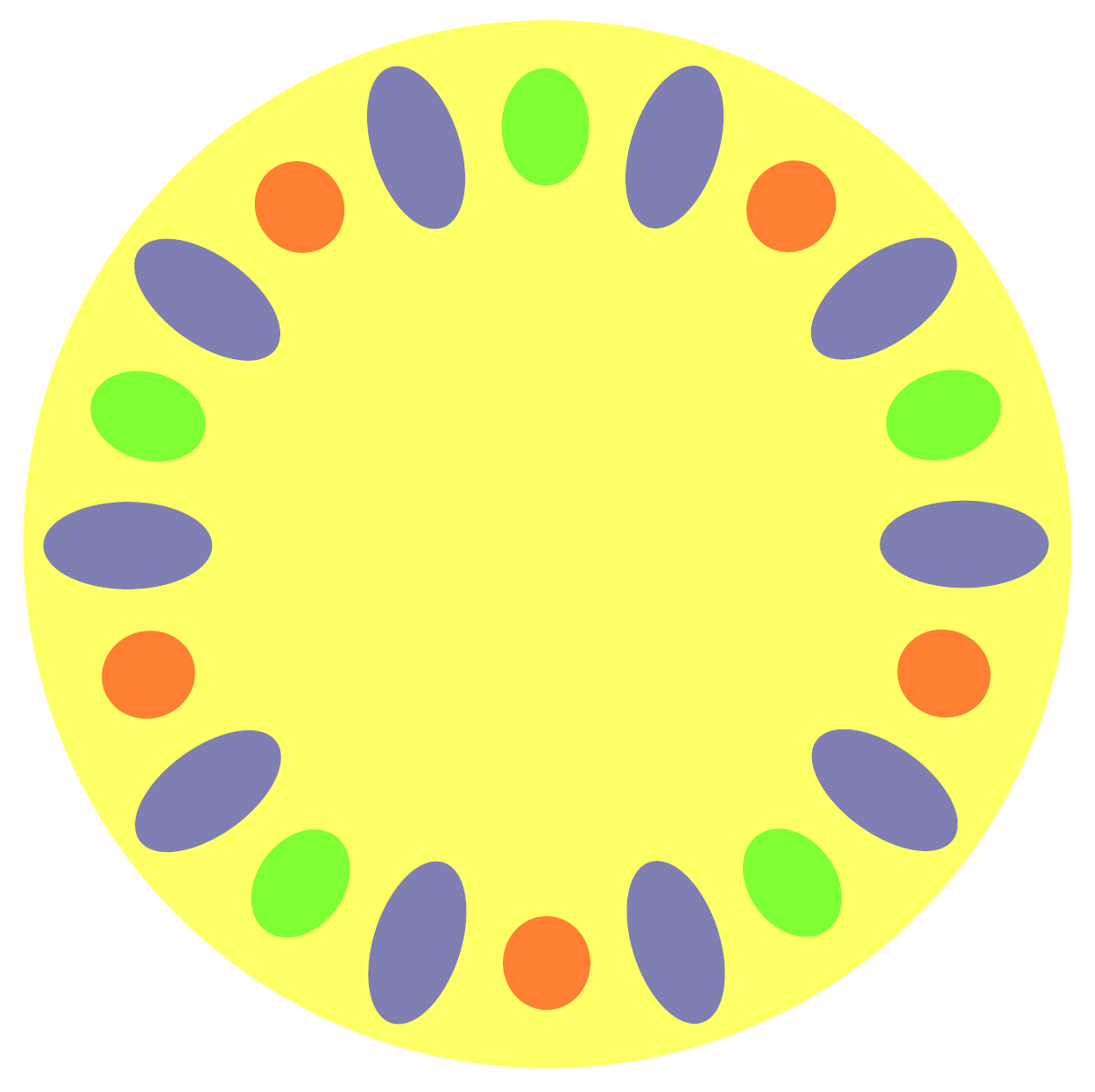}

}

\caption{The illuminated area on the target by 20 OMEGA beams. The transverse
section of the beams are circles, but the spots on the target surface
are ellipses due to inclination. The incident angle is 59$\lyxmathsym{\protect\textdegree}$
for \textcolor{blue}{blue} spot, 42$\lyxmathsym{\protect\textdegree}$
for \textcolor{green}{green} spot, and 21$\lyxmathsym{\protect\textdegree}$
for \textcolor{red}{red} spot. (a) $d=400\mathrm{\mu m}$ ring radius;
(b) $d=800\mathrm{\mu m}$ ring radius; (c) $d=1200\mathrm{\mu m}$
ring radius. The $d=0$ case is not shown. Note that the red and green
spots form a 5-fold symmetry and the blue spots form a 10-fold symmetry.\label{fig:The-illuminated-area}}
\end{figure}

\begin{figure}[th]
\hfill{}\includegraphics[scale=0.35]{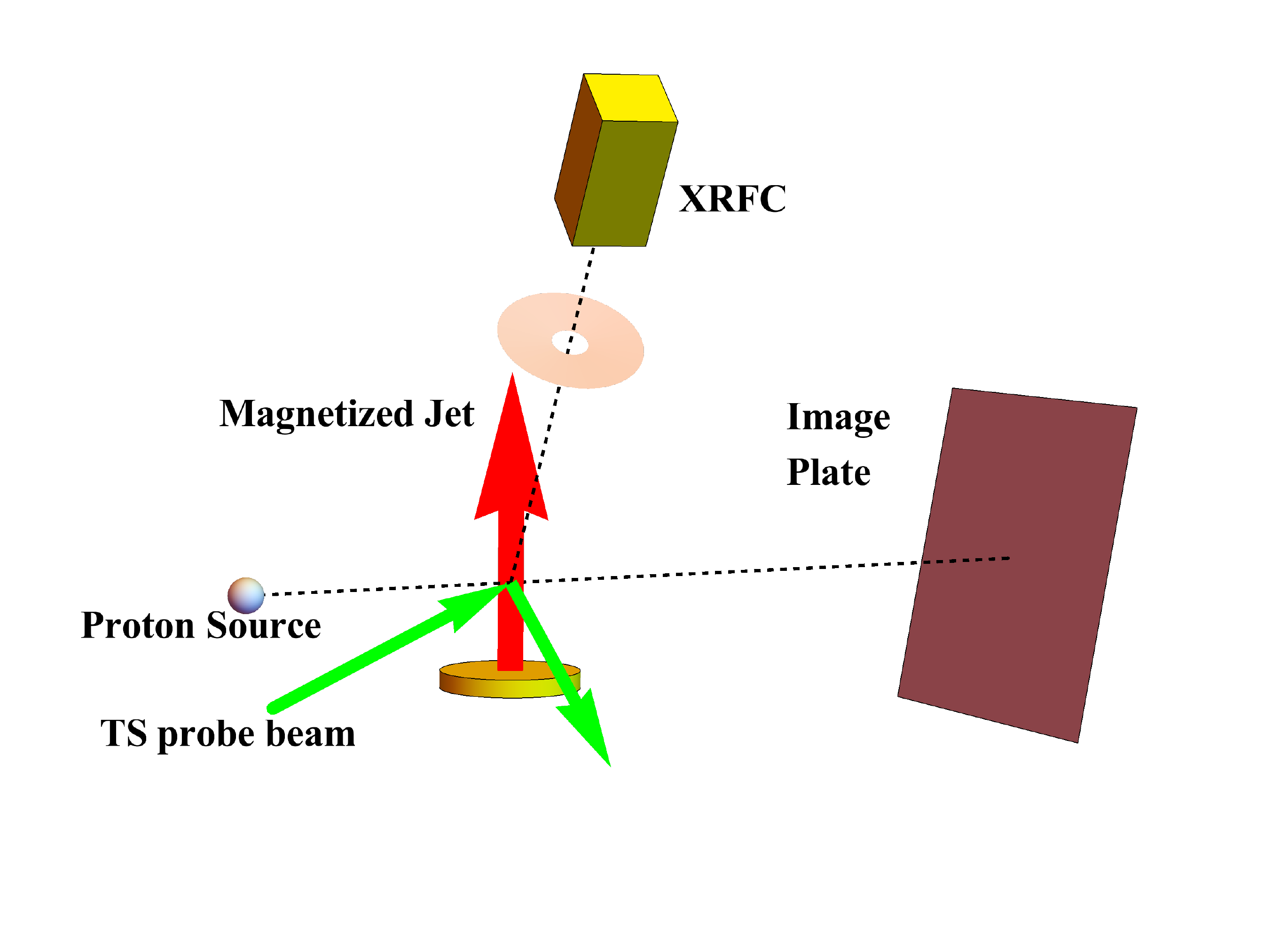}\hfill{}

\caption{Schematics of diagnostics setup of OMEGA magnetized jet experiments.
For DD and $\mathrm{D^{3}He}$ protons, the source stands 1cm from
TCC, while the image plate CR39 is located 17cm from TCC on the other
side. For the TNSA protons, the source stands 0.8cm from TCC, while
the radiochromic film pack is located 16.5cm from TCC on the other
size. The X-ray framing camera(XRFC) images the jet from at 38$\lyxmathsym{\protect\textdegree}$
from axis, which is not modeled in this work.\label{fig:Schematics-of-diagnostics}}
\end{figure}

To model the laser driven blowoffs, we use the spatial and temporal
specifications of each of the twenty OMEGA driver beams. The 20 driver
beams are turned on and turned off simultaneously with a 1ns pulse
duration. Each delivers 500J of energy on a target flat-top. The radius
of each beam is 125$\mathrm{\mu m}$. The laser spots are arranged
to form a ring pattern of radius $d$, as shown in Fig \ref{fig:The-illuminated-area}.
The target is 0.5mm thick to prevent the burn-through. The setup of
the diagnostics is sketched in Figure \ref{fig:Schematics-of-diagnostics}
and discussed in the following subsections.

For convention, $t=0$ is the time for laser turn on. The $z$ direction
is perpendicular to the surface of the target plane. The jet is formed
in the $z>0$ region. We also use cylindrical coordinates where $r=0$
is the central axis of the target. The target surface is located at
$z=0$. Axial direction is along $z$ axis. Toroidal or azimuthal
direction is the $\varphi$ direction in the cylindrical coordinate
system. Target chamber center(TCC) is at $x=y=r=0$, $z=0.25\mathrm{cm}$.

\subsection{Optical Thomson scattering}

Optical Thomson scattering\citep{Froula2006,Froula2005} is used to
probe the electron/ion temperatures, electron density and flow velocity
at TCC. We used one probe beam with 1ns pulse, $25\sim50\mathrm{J}$
energy and 532nm wavelength($2\omega$) as the backlighter. The intensity
distribution of the probe beam is 70$\mathrm{\mu m}$ FWHM 2D gaussian. 

We model the Thomson scattering spectrum using the 3D FLASH simulation
results. The spatial profiles of electron density, electron/ion temperature,
flow velocity, and fraction of species are taken as the input for
the spectroscopy code\citep{TS_Sheffield2010}. The heating by the
probe beam is modeled by laser absorption. The dispersion relations
for ion acoustic wave and electron plasma wave\citep{TS_Sheffield2010}
are used to calculate the power output. The final power output is
weight averaged by the spatial intensity distribution of the probe
laser. The instrument broadening\citep{TS_Follett2016} is taken into
account in the modeling spectrum. 

The experimental data are also fitted using the model(see more results
in Gao et. al. 2018) to compare with the plasma quantities averaged
$(200\mathrm{\mu m})^{3}$ cube centered at TCC. 

\subsection{Proton radiography}

Our OMEGA experiments used two types of proton sources to map out
the magnetic fields (1) DD(3MeV) and $\mathrm{D^{3}He}$(14.7MeV)
protons from fusion reaction driven by 24 OMEGA beams\citep{Prad_Li2006,Prad_Manuel2012}.
The actual spectrum is typically an up-shifted symmetric gaussian
distribution, FWHM=320keV centered at 3.6MeV for DD protons, FWHM=670keV
centered at 15.3MeV for $\mathrm{D^{3}He}$ protons. The emitting
position of protons follows a 3D gaussian distribution with e-fold
radius equal to 20$\mu\mathrm{m},$ and the burn time is 150ps\citep{Prad_Manuel2012};
(2) Broadband protons up to 60MeV are driven by an OMEGA EP beam via
Target Normal Sheath Acceleration(TNSA) mechanism\citep{Prad_Zylstra2012}.
The actual spectrum is typically an exponential distribution with
effective temperature 3.79MeV for our copper backlighter target\citep{Prad_Flippo2010}.
The initial position of protons at the source follows a 3D gaussian
distribution with e-fold radius equal to 5$\mu\mathrm{m}$\citep{Prad_Manuel2012},
and the pulse duration is 1ps. For the DD and $\mathrm{D^{3}He}$
protons, the source stands 1cm from TCC, while the image plate CR39
is located 17cm from TCC on the other side. For the TNSA protons,
the source stands 0.8cm from TCC, while the radiochromic film pack
is located 16.5cm from TCC on the other side. 

The modeling for proton radiography is composed of (1) sampling for
the source distributions mentioned above, (2) solving the trajectory
of the protons, (3) recording the protons on the detector plane.

The deflection of protons in electromagnetic fields is calculated
by solving the Newton-Lorentz equation
\begin{equation}
\frac{d(m_{p}\boldsymbol{v})}{dt}=e(\boldsymbol{E}+\frac{\boldsymbol{v}}{c}\times\boldsymbol{B})
\end{equation}
In a typical MHD fluid, $E\approx\frac{v_{h}}{c}B$, where $v_{h}$
is the hydrodynamical velocity scale of the fluid. The ratio of electric
force to magnetic force is $\frac{E}{(v_{p}/c)B}\approx\frac{v_{h}}{v_{p}}$.
For a proton with energy larger than $\sim\mathrm{MeV}$, the proton
speed $v_{p}$ is much larger than $v_{h}$, so we use electric field
$\boldsymbol{E}=0$ approximation in the modeling. The energy lost
is calculated throughout the proton motion from the NIST PSTAR table\footnote{The PSTAR table is available at \href{https://physics.nist.gov/PhysRefData/Star/Text/PSTAR.html}{https://physics.nist.gov/PhysRefData/Star/Text/PSTAR.html}}.
Protons lose significant amount of energy in the remaining solid target
with density $\sim1\mathrm{g/cc}$.

We assume that the detector for DD protons has uniform sensitivity
for protons with $E>2\mathrm{MeV}$, and that for $\mathrm{D^{3}He}$
protons has uniform sensitivity for all protons with $E>14.4\mathrm{MeV}$.
The TNSA proton energy range that each film is primarily sensitive
to is $E>E_{0}$, and the deposited energy per proton is proportional
to $(E-E_{0})^{-1/2}$, while no energy is deposited for $E<E_{0}$.
The characteristic energy $E_{0}$ is different for each pack in radiochromic
film. Temporal smearing of TNSA protons images is neglected because
the pulse duration is short. Temporal smearing of DD and $\mathrm{D^{3}He}$
protons is calculated using the integral of the second order interpolation
among successive images with 0.1ns intervals. 

\section{\label{sec:FLASH-Simulation-results}FLASH Simulation results}

\subsection{Hydrodynamics}

\begin{figure}
\hfill{}\subfloat[]{\includegraphics[scale=0.12]{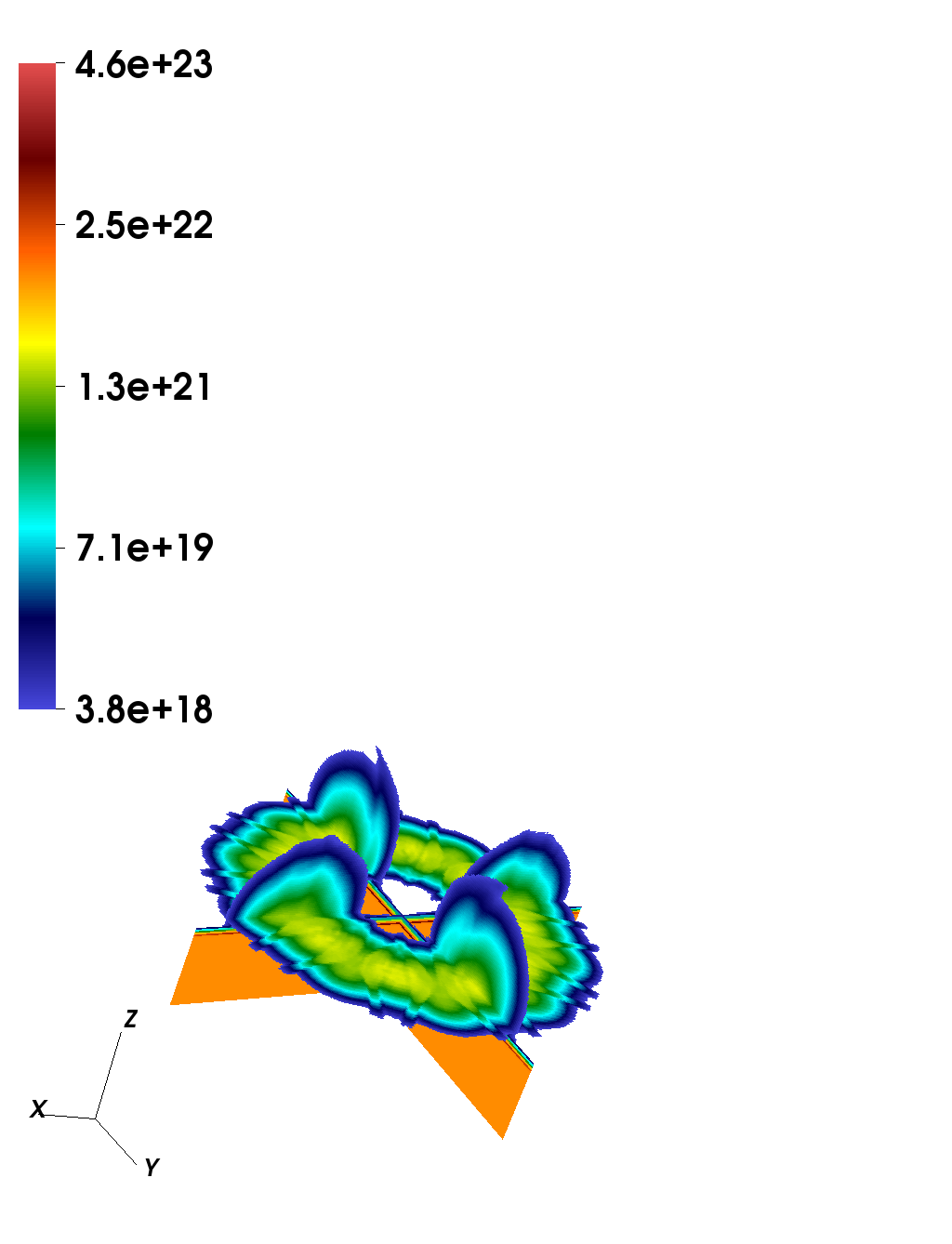}}\subfloat[]{\includegraphics[scale=0.12]{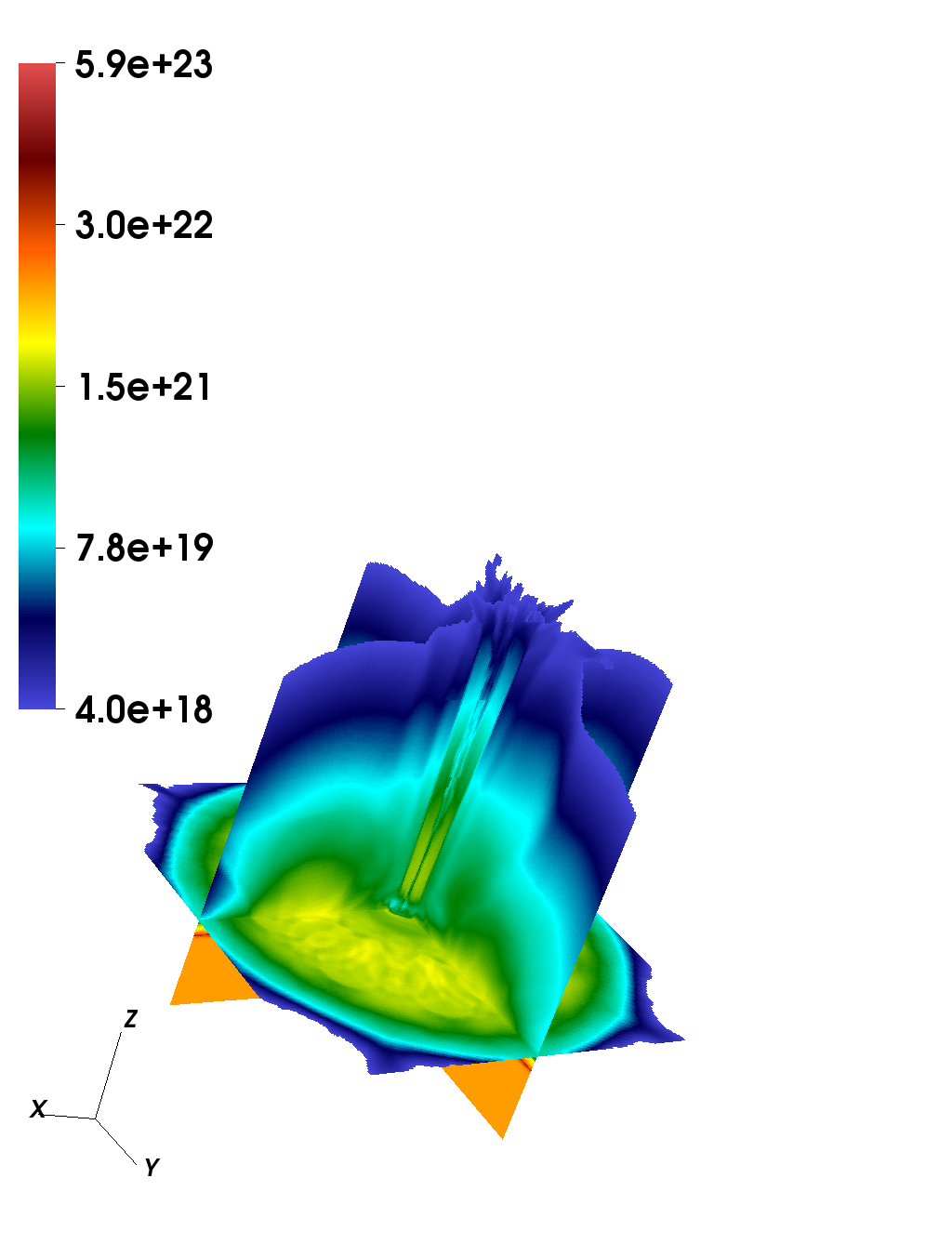}}\hfill{}

\hfill{}\subfloat[]{\includegraphics[scale=0.12]{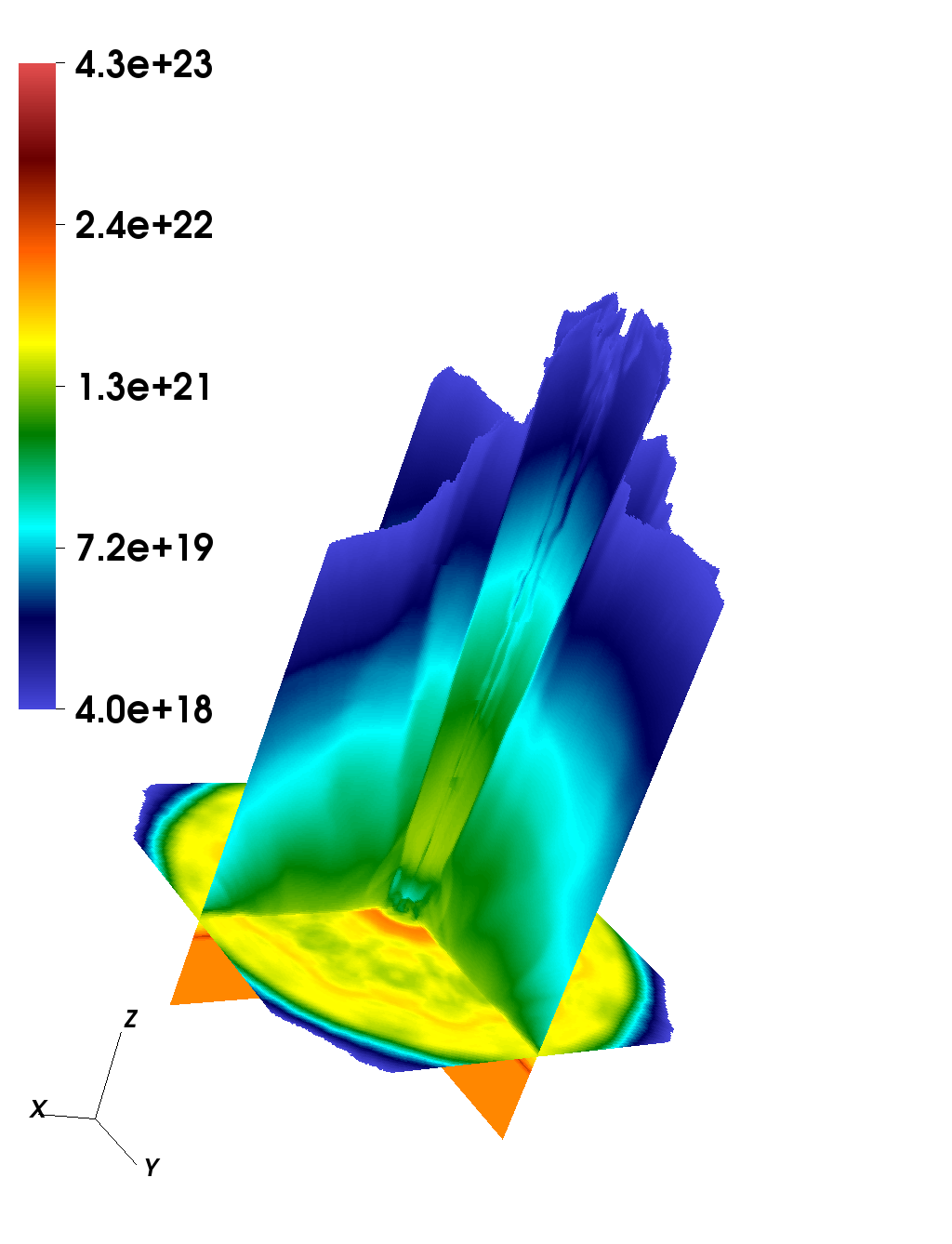}}\subfloat[]{\includegraphics[scale=0.12]{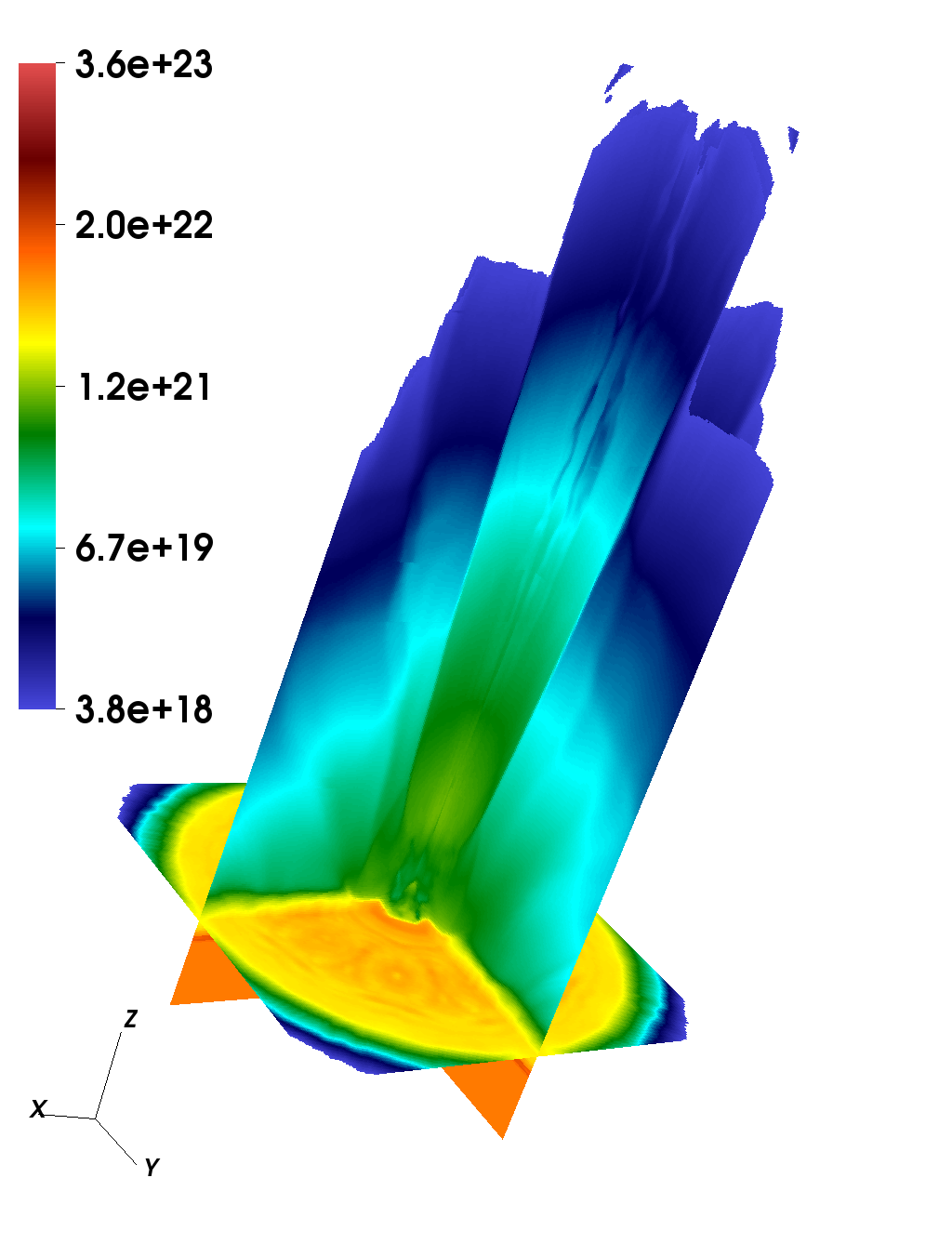}

}\hfill{}

\caption{Three-slice plots for electron density at $x=0$, $y=0$ and $z=0.01\mathrm{cm}$
planes in FLASH simulations. The unit is $\mathrm{cm^{-3}}$. The
$z=0.01\mathrm{cm}$ plane in the simulation is $(0.3\mathrm{cm)^{2}}$
rectangle. The ring radius is $d=800\mathrm{\mu m}$, and the target
is CH. (a) $t=0.6\mathrm{ns}$ (when laser is still on), (b) $t=1.4\mathrm{ns}$($0.4\mathrm{ns}$
after the laser is turned off) (c) $t=2.2\mathrm{ns}$, (d) $t=3.0\mathrm{ns}$.
These plots show the time evolution of the jet.\label{fig:nele_dt}}
\end{figure}

The jet is formed by the merging of the plasma plumes produced by
20 individual OMEGA beams through a strong cylindrical shock. By using
a large ring radius, the flows will not collide immediately while
the lasers irradiate the target. For the collision at later time with
more available room, the flows develop larger radial velocities which
become more supersonic. Thus a stronger cylindrical shock is generated
near the $z$ axis. For the cylindrical shock, the surrounding is
in the upstream and the central core is in the downstream. It is a
hydrodynamic shock where the plasma $\beta$ is much larger than unity.
Figure \ref{fig:nele_dt} shows the time evolution of the jet for
$d=800\mathrm{\mu m}$ CH target simulation. The jet is supersonic
and well collimated. The jets with different ring radius all travel
several millimeters by $t=\mathrm{3ns}$. The jet keeps traveling
and expanding so that the length $L$ and the radius $R$ keep growing
even after $3\mathrm{ns}$. The width and the length of the jet are
much larger than the laser spot size(250$\mathrm{\mu m}$), as shown
in Table \ref{tab:Comparison-of-plasma}.

\begin{figure}
\hfill{}\subfloat[]{\includegraphics[scale=0.12]{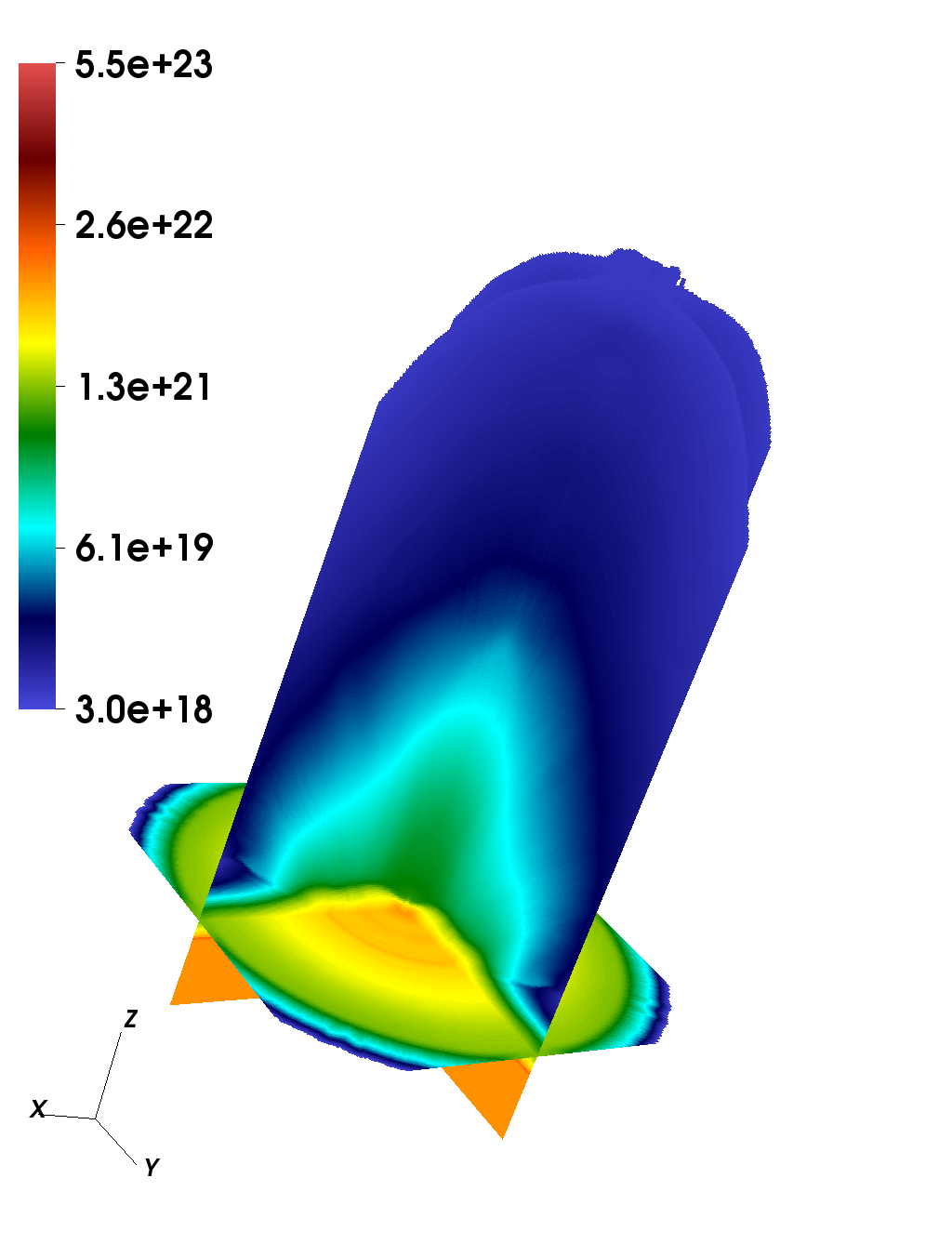}}\subfloat[]{\includegraphics[scale=0.12]{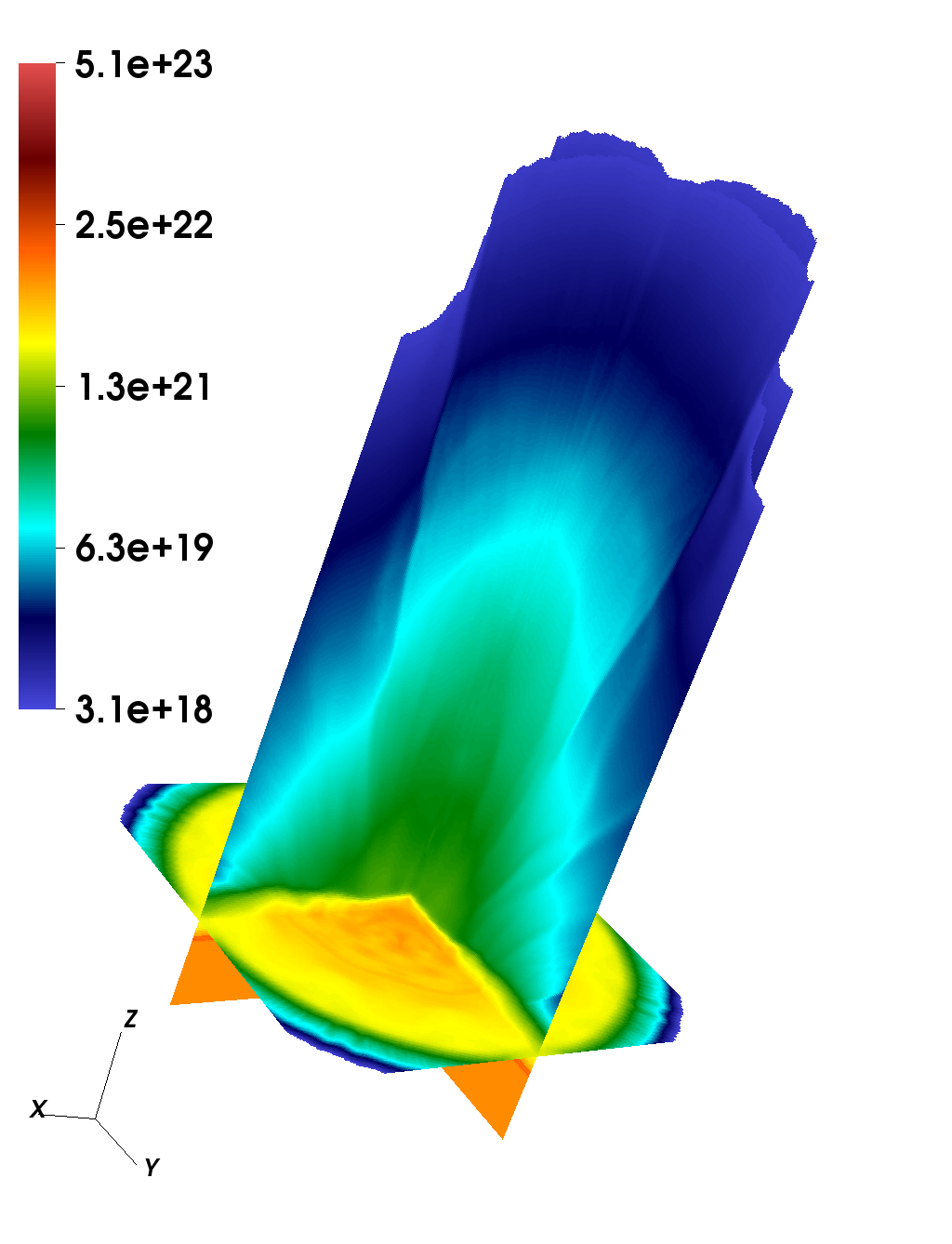}}\hfill{}

\hfill{}\subfloat[]{\includegraphics[scale=0.12]{figures_hd/nele_3pt0ns}

}\subfloat[]{\includegraphics[scale=0.12]{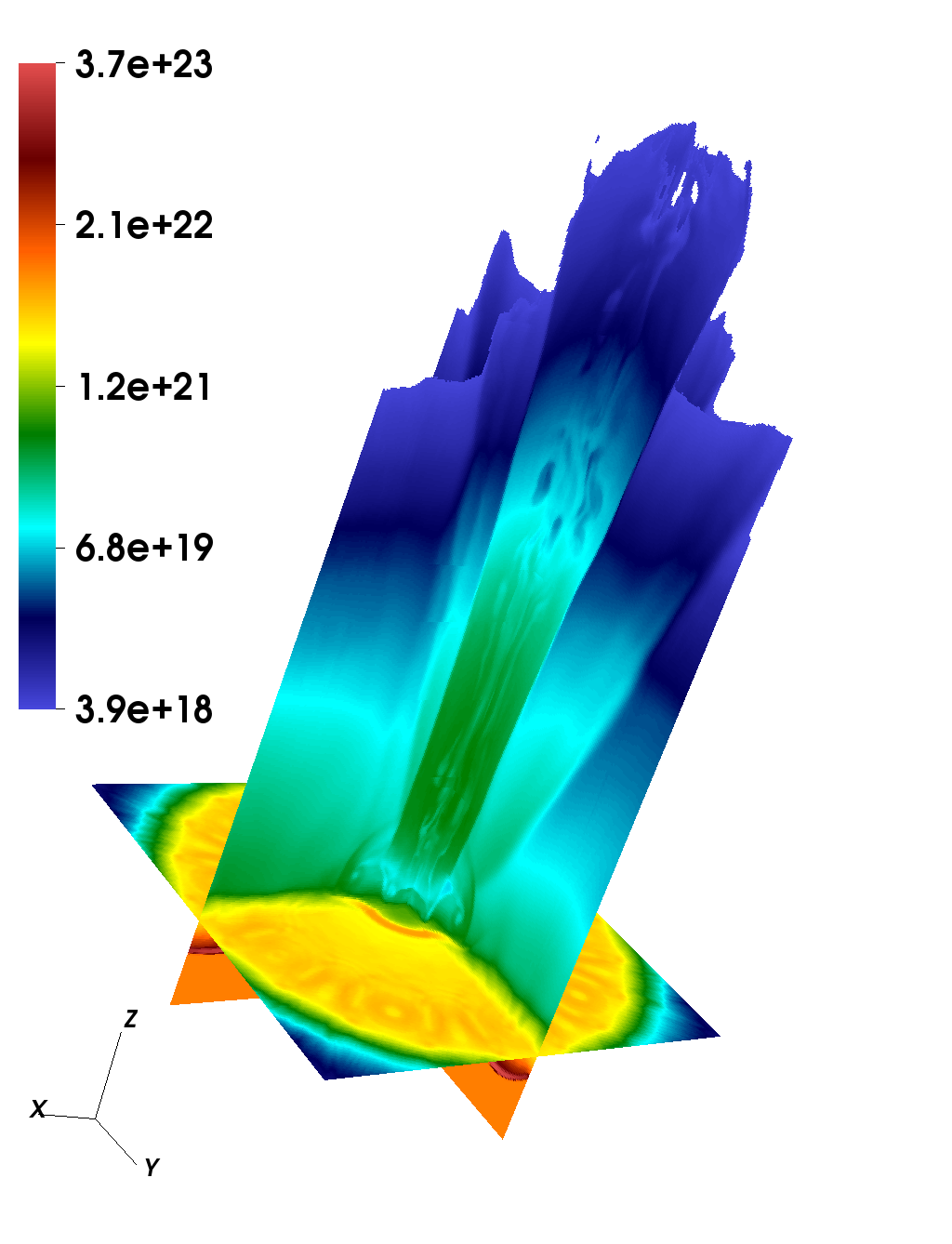}}\hfill{}\caption{Three-slice plots for electron density at $x=0$, $y=0$ and $z=0.01\mathrm{cm}$
planes at $t=3\mathrm{ns}$ for four different ring radii $d$ in
FLASH simulations. The unit is $\mathrm{cm^{-3}}$. The scale is as
same as Figure \ref{fig:nele_dt}. The targets are CH. (a) $d=0$,
(b) $d=400\mathrm{\mu m}$, (c) $d=800\mathrm{\mu m}$, (d) $d=1200\mathrm{\mu m}$.
Figure (c) here is as same as Figure \ref{fig:nele_dt}(d). \label{fig:nele_dd}}
\end{figure}

\begin{figure}
\includegraphics[scale=0.6]{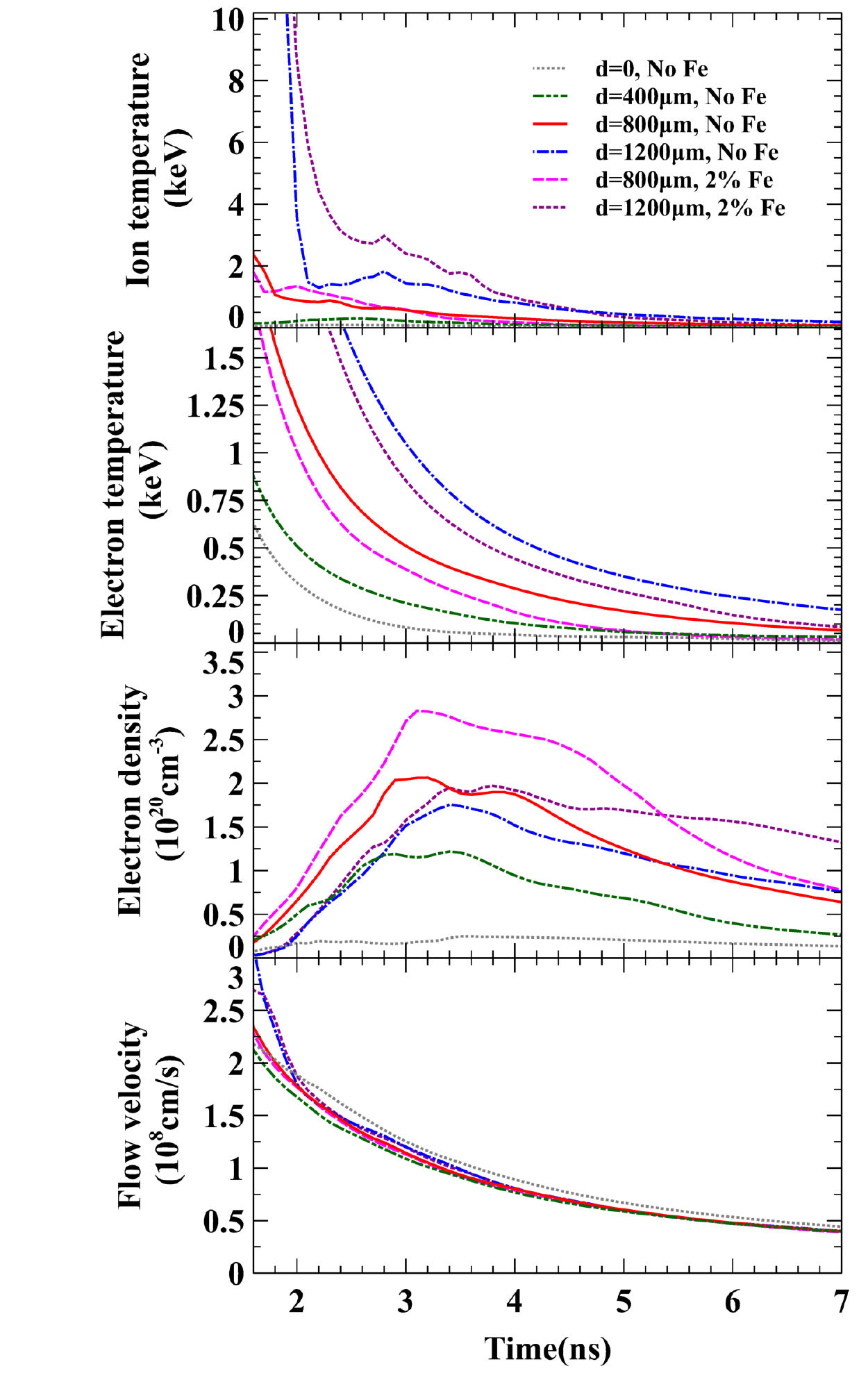}

\caption{The evolution of plasma variables at TCC for the six different runs
in FLASH simulation. Four sub-figures share the time axis and legends.
The quantities are calculated by averaging over a $(200\mathrm{\mu m})^{3}$
cubic around TCC. \label{fig:plasma_tcc}}
\end{figure}

\begin{table*}[tp]
\caption{Comparison of plasma properties for different ring radii and targets
at $t=3\mathrm{ns}$ and $r=0$, $z=2.5\mathrm{mm}$. The $n_{e}$,
$\rho$, $T_{e}$, $T_{i}$ and $B$ are calculated by averaging over
a $(200\mathrm{\mu m)^{3}}$ cubic around TCC. The jet length $L$
is the defined by the point on the $z$ axis where electron density
drops to $3\times10^{18}\mathrm{cm}^{-3}$. The radius $R$ is defined
by reading the position in $z=2.5\mathrm{mm}$ plane where the density
scale height $|\nabla\log\rho|^{-1}$ reaches minimum. Columns 2 to
5 are for pure CH targets. Columns 6 and 7 are for 2\% Fe dopant targets.\label{tab:Comparison-of-plasma}}

\begin{tabular}{|c|c|c|c|c|c|c|}
\hline 
Plasma property & $d=0$ & $d=400\mathrm{\mu m}$ & $d=800\mathrm{\mu m}$ & $d=1200\mathrm{\mu m}$ & \makecell{$d=800\mathrm{\mu m}$\\2\% Fe dopant} & \makecell{$d=1200\mathrm{\mu m}$\\2\% Fe dopant}\tabularnewline
\hline 
\hline 
Electron density $n_{e}$($\mathrm{cm}^{-3}$) & $1.7\times10^{19}$ & $1.2\times10^{20}$ & $2.0\times10^{20}$ & $1.5\times10^{20}$ & $2.7\times10^{20}$ & $1.6\times10^{20}$\tabularnewline
\hline 
Electron temperature $T_{\mathrm{e}}$(eV) & $81$ & $2.1\times10^{2}$ & $5.1\times10^{2}$ & $1.0\times10^{3}$ & $3.9\times10^{2}$ & $8.5\times10^{2}$\tabularnewline
\hline 
Ion temperature $T_{\mathrm{i}}$(eV) & $76$ & $2.2\times10^{2}$ & $5.7\times10^{2}$ & $1.4\times10^{3}$ & $5.9\times10^{2}$ & $2.4\times10^{3}$\tabularnewline
\hline 
Magnetic field $B$ (gauss) & $2.4\times10^{4}$ & $1.4\times10^{5}$ & $3.3\times10^{5}$ & $3.1\times10^{5}$ & $3.5\times10^{5}$ & $3.7\times10^{5}$\tabularnewline
\hline 
Jet width $R$(cm) & $>0.15$ & $0.091$ & $0.049$ & $0.039$ & $0.047$ & $0.042$\tabularnewline
\hline 
Jet length $L$(cm) & $0.46$ & $0.52$ & $0.54$ & $0.53$ & $0.55$ & $0.50$\tabularnewline
\hline 
$L/R$ & $<3.1$ & $5.7$ & $11$ & $13.6$ & $11.7$ & $11.9$\tabularnewline
\hline 
\end{tabular}
\end{table*}

The properties of the jet become more interesting as the ring radius
$d$ increases. Figure \ref{fig:nele_dd} shows the shape of the jet
for different laser ring radii at $t=3\mathrm{ns}$. Figure \ref{fig:plasma_tcc}
shows the evolution of electron/ion temperature, electron density
and flow velocity at TCC for different runs in the FLASH simulation.
The quantities are calculated by averaging over a $(200\mathrm{\mu m)^{3}}$
cubic around TCC($r=0$, $z=0.25\mathrm{cm}$). The peak electron/ion
temperature on-axis is higher for larger ring radius. Comparing to
the case where $d=0$, the temperatures for $d=800\mathrm{\mu m}$
or $d=1200\mathrm{\mu m}$ are about one order of magnitude higher.
The peak electron density is highest for $d=800\mathrm{\mu m}$, which
is one order of magnitude higher than the $d=0$ case. The ratio $L/R$
of the jet becomes larger(see Table \ref{tab:Comparison-of-plasma})
as $d$ gets larger. Large ring radius also reduces the opening angle
of the jet. The flow velocity is hardly affected by increasing $d$.
These results are in good agreement with previous 2D cylindrical hydrodynamics
simulations by Fu et. al.\citep{Fu_Fu2013} using 2D FLASH. The simulations
in this work are in 3D cartesian geometry. The full details of the
laser configuration are taken into account. In 3D simulations, even
though there is the azimuthal asymmetry of the laser intensity on
the target as shown in Figure \ref{fig:The-illuminated-area}, the
jet is still well collimated and has similar hydrodynamical properties
as in the 2D cylindrical case. The azimuthal asymmetry level for electron
density can exceed 10\%, and the pattern of density distribution in
z-slice resembles a ``sun flower'' as shown in Figure \ref{fig:Slice-at-tcc-b}.

\begin{figure}
\hfill{}\subfloat[]{\includegraphics[scale=0.12]{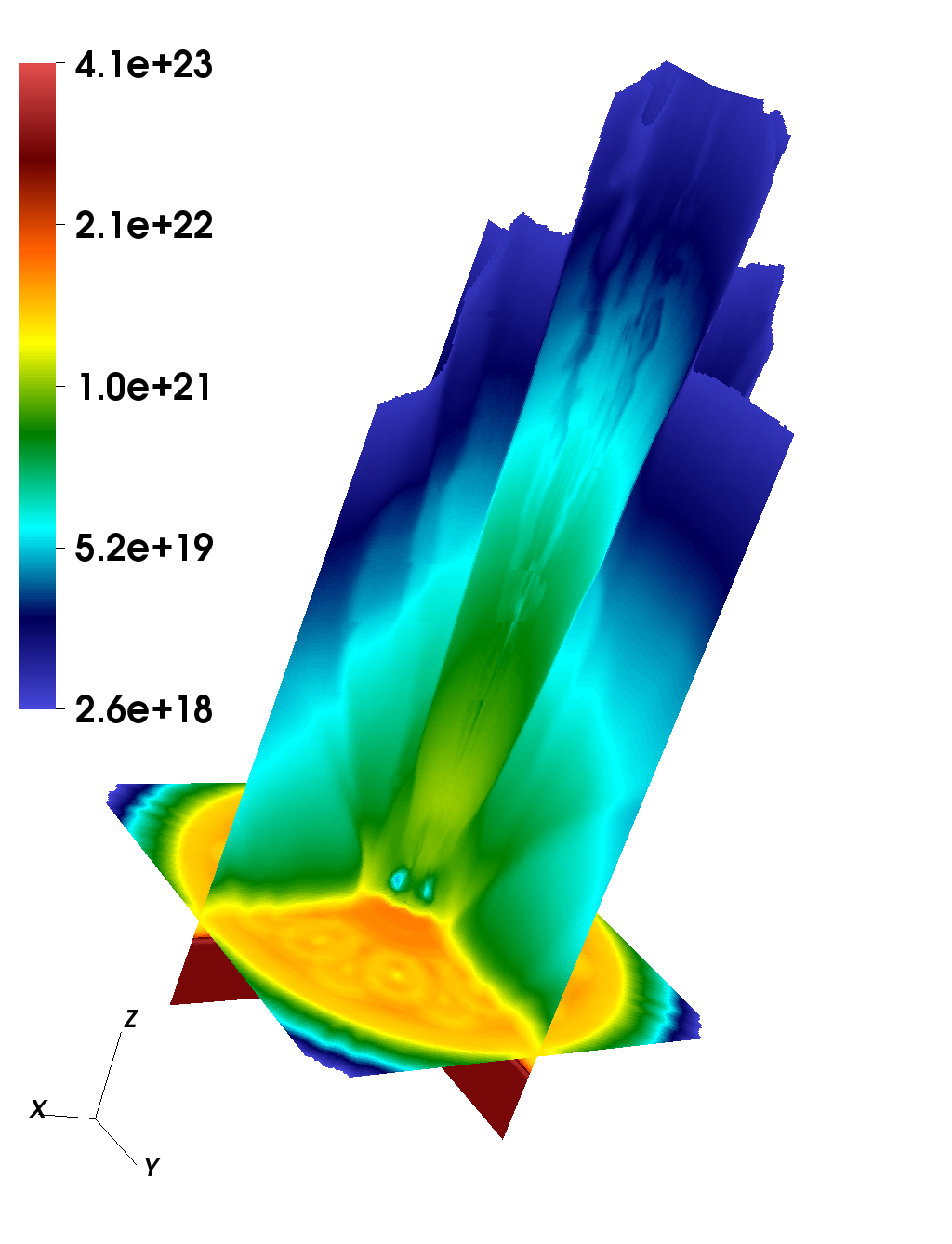}}\subfloat[]{\includegraphics[scale=0.12]{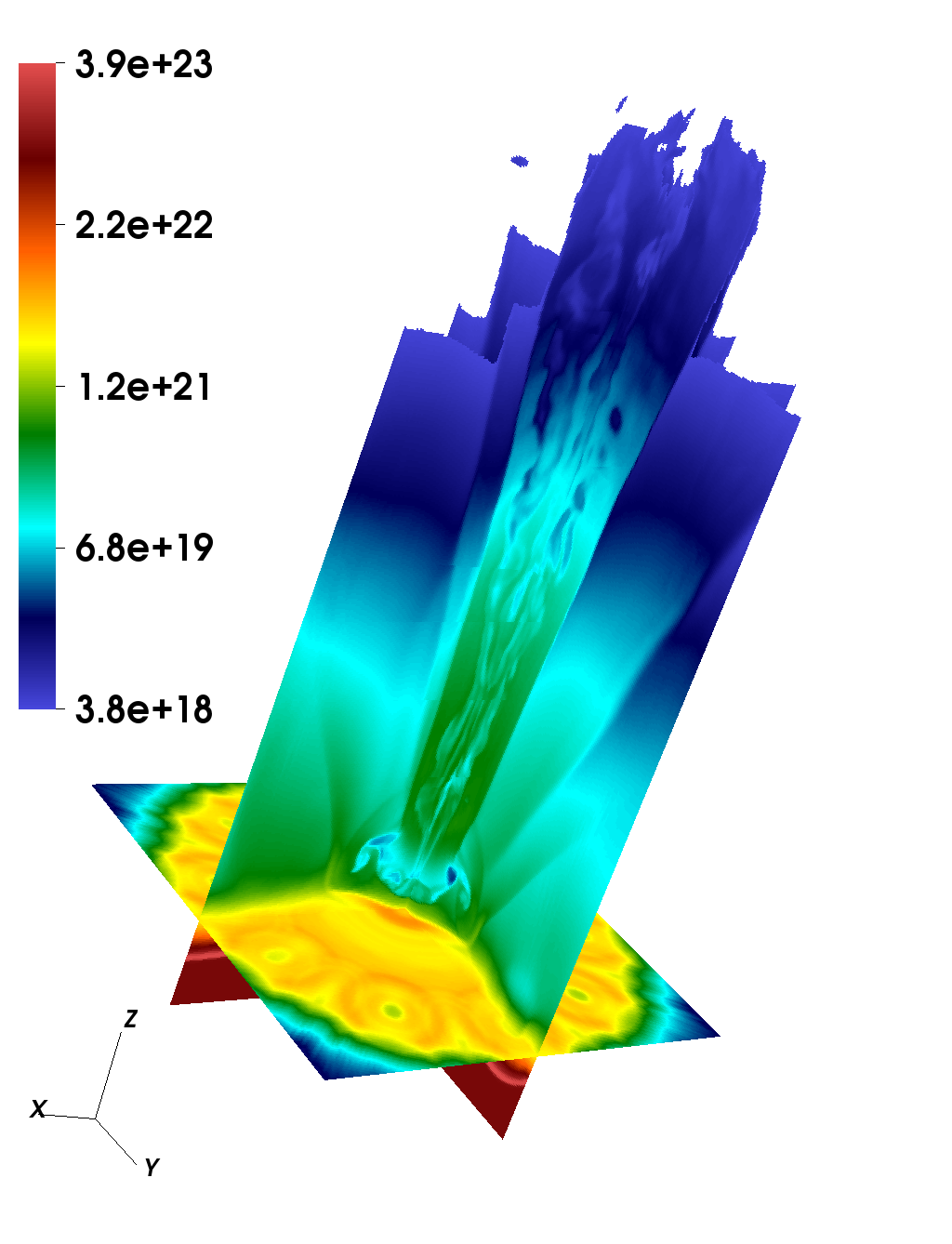}}\hfill{}

\hfill{}\subfloat[]{\includegraphics[scale=0.12]{figures_hd/nele_3pt0ns}

}\subfloat[]{\includegraphics[scale=0.12]{figures_hd/nele_1200d}}\hfill{}

\caption{Three-slice plots for electron density(unit: $\mathrm{cm^{-3}}$)
at $x=0$, $y=0$ and $z=0.01\mathrm{cm}$ planes at $t=3\mathrm{ns}$
for two different ring radii $d$ and for two different types of targets
in FLASH simulations. The unit is $\mathrm{cm^{-3}}$. The scale is
as same as Figure \ref{fig:nele_dt}. (a) $d=800\mathrm{\mu m}$,
2\% Fe-doped target, (b) $d=1200\mathrm{\mu m}$, 2\% Fe-doped target
(c) $d=800\mathrm{\mu m}$, CH target, (d) $d=1200\mathrm{\mu m}$,
CH target. Figure (c) here is as same as Figure \ref{fig:nele_dd}(c),
and Figure (d) here is as same as Figure \ref{fig:nele_dd}(d). \label{fig:nele_ddop}}
\end{figure}

The jet for 2\% Fe dopant shot is slightly different than the one
without dopant, as shown in Figure \ref{fig:nele_ddop}. The jet in
a dopant shot radiate several times more than that in a non-dopant
shot. But the radiative cooling time at $t=\mathrm{3ns}$ for the
jet is much large than nanosecond even in the dopant shot. Thus, the
radiation cooling(see Table \ref{tab:Radiation-properties-of}) has
little to do with the shape of the jet after it has grown to millimeter
size. For an earlier time, however, cooling rate is large enough to
play a role. As a result, the electron temperature at TCC for doped
jets is always lower than that in the non-doped jets with the same
ring radius $d$. The reduction in electron temperature relaxes the
cylindrical shock. Thus, more electrons flow into the core, which
causes the jets in dopant shots to have higher electron density than
the non-dopant ones. In both doped and non-doped case, the jets are
always optically thin.

\begin{table*}
\caption{Simulated plasma properties for case $d=800\mu$m, $t=3$ns at $z=2.5\mathrm{mm}$
in non-dopant run. All quantities are in cgs units except temperatures
expressed in eV. The length scale is $L$ approximately the width
of the jet at $z=2.5\mathrm{mm}$, which is $L\approx1\mathrm{mm}$.
The $n_{e}$, $\rho$, $T_{e}$ and $T_{i}$ at $r=0$ are calculated
by averaging over a $(200^{3}\mathrm{\mu m})^{3}$ cubic around TCC.
The $n_{e}$, $\rho$, $T_{e}$ and $T_{i}$ at $r=1\mathrm{mm}$
are calculated by averaging over a $200\mathrm{\mu m}$ high and $200\mathrm{\mu m}$
thick ring around $r=1\mathrm{mm}$, $z=2.5\mathrm{mm}$. $B$ is
calculated using the root of mean square in the same cubic. The variation
of $n_{e}$ is $\Delta n_{e}=\sqrt{\overline{n_{e}^{2}}-\overline{n_{e}}^{2}}$,
similar for other variables. \label{tab:Simulated-plasma-properties}}

\begin{tabular}{|c|c|c|c|}
\hline 
Plasma property & Formula & Value at $r=0$ & Value at $r=1\mathrm{mm}$\tabularnewline
\hline 
\hline 
Electron density $n_{e}$($\mathrm{cm}^{-3}$) & $\cdots$ & $2.0\times10^{20}$ & $3.0\times10^{19}$\tabularnewline
\hline 
$\Delta n_{e}$($\mathrm{cm}^{-3}$) & $\cdots$ & $1.5\times10^{19}$ & $4.1\times10^{18}$\tabularnewline
\hline 
Mass density $\rho$($\mathrm{g/cm^{3}}$) & $\cdots$ & $6.3\times10^{-4}$ & $1.1\times10^{-4}$\tabularnewline
\hline 
$\Delta\rho$($\mathrm{g/cm^{3}}$) & $\cdots$ & $4.7\times10^{-5}$ & $1.3\times10^{-5}$\tabularnewline
\hline 
Electron temperature $T_{\mathrm{e}}$(eV) & $\cdots$ & $5.1\times10^{2}$ & $3.9\times10^{2}$\tabularnewline
\hline 
 $\Delta T_{\mathrm{e}}$(eV) & $\cdots$ & $1.2$ & $1.4$\tabularnewline
\hline 
Ion temperature $T_{\mathrm{i}}$(eV) & $\cdots$ & $5.7\times10^{2}$ & $1.8\times10^{2}$\tabularnewline
\hline 
 $\Delta T_{\mathrm{i}}$(eV) & $\cdots$ & $1.2\times10^{2}$ & $17$\tabularnewline
\hline 
Magnetic field $B$ (gauss) & $\cdots$ & $3.3\times10^{5}$ & $4.6\times10^{4}$\tabularnewline
\hline 
$\Delta B$ (gauss) & $\cdots$ & $1.5\times10^{5}$ & $3.0\times10^{4}$\tabularnewline
\hline 
Average ionization $Z$ & $\cdots$ & $3.5$ & $3.5$\tabularnewline
\hline 
Average atomic weight $A$ & $\cdots$ & $6.5$ & $6.5$\tabularnewline
\hline 
Flow velocity $u\approx u_{z}$(cm/s) & $\cdots$ & $1.1\times10^{8}$ & $1.1\times10^{8}$\tabularnewline
\hline 
$\Delta u$(cm/s) & $\cdots$ & $3.2\times10^{6}$ & \tabularnewline
\hline 
Perpendicular velocity $\sqrt{u_{x}^{2}+u_{y}^{2}}$(cm/s) & $\cdots$ & $2.7\times10^{6}$ & $1.5\times10^{7}$\tabularnewline
\hline 
Sound speed $c_{s}$(cm/s) & $9.8\times10^{5}$$\frac{[ZT_{\mathrm{ele}}+1.67T_{\mathrm{ion}}]^{1/2}}{A^{1/2}}$ & $2.0\times10^{7}$ & $1.6\times10^{7}$\tabularnewline
\hline 
Mach number $M$ & $u/c_{s}$ & $5.5$ & $6.8$\tabularnewline
\hline 
Electron plasma frequency(rad/s) & $5.6\times10^{4}n_{e}^{1/2}$ & $7.9\times10^{14}$ & $3.1\times10^{14}$\tabularnewline
\hline 
Coulomb logarithm $\ln\Lambda$ & \makecell{$23.5-\ln(n_{e}^{1/2}T_{e}^{-5/4})-$ \\ $[10^{-5}+(\ln T_{e}-2)^{2}/16]^{1/2}$} & $6.9$ & $7.5$\tabularnewline
\hline 
Electron thermal velocity $v_{Te}$(cm/s) & $4.2\times10^{7}T_{e}^{1/2}$ & $9.5\times10^{8}$ & $8.3\times10^{8}$\tabularnewline
\hline 
Electron collision rate $\nu_{e}$(1/s) & $2.9\times10^{-6}n_{e}\ln\Lambda T_{e}^{-3/2}$ & $3.5\times10^{11}$ & $8.5\times10^{10}$\tabularnewline
\hline 
Electron-ion collision rate $\nu_{ei}$(1/s) & $3.2\times10^{-9}ZA^{-1}n_{e}\ln\Lambda T_{e}^{-3/2}$ & $2.1\times10^{8}$ & $5.1\times10^{7}$\tabularnewline
\hline 
Electron mean free path $l_{e}$(cm) & $v_{Te}/\nu_{e}$ & $2.7\times10^{-3}$ & $9.8\times10^{-3}$\tabularnewline
\hline 
Electron gyro-frequency $\omega_{ce}$(rad/s) & $1.7\times10^{7}B$ & $5.6\times10^{12}$ & $7.8\times10^{11}$\tabularnewline
\hline 
Electron gyroradius $r_{e}$(cm) & $2.4T_{e}^{1/2}B^{-1}$ & $1.6\times10^{-4}$ & $7.0\times10^{-4}$\tabularnewline
\hline 
Ion thermal velocity $v_{Ti}$(cm/s) & $9.8\times10^{5}A^{-1/2}T_{i}^{1/2}$ & $9.2\times10^{6}$ & $2.9\times10^{6}$\tabularnewline
\hline 
Ion collision rate $\nu_{i}$(1/s) & $4.8\times10^{-8}Z^{3}n_{e}A^{-1/2}\ln\Lambda T_{i}^{-3/2}$ & $8.2\times10^{10}$ & $1.8\times10^{14}$\tabularnewline
\hline 
Ion mean free path $l_{i}$(cm) & $v_{Ti}/\nu_{i}$ & $1.1\times10^{-4}$ & $1.6\times10^{-8}$\tabularnewline
\hline 
Ion gyro-frequency $\omega_{ci}$(rad/s) & $9.6\times10^{3}ZB/A$ & $1.7\times10^{9}$ & $2.4\times10^{8}$\tabularnewline
\hline 
Ion gyroradius $r_{i}$(cm) & $1.0\times10^{2}A^{1/2}Z^{-1}T_{i}^{1/2}B^{-1}$ & $5.3\times10^{-3}$ & $2.1\times10^{-2}$\tabularnewline
\hline 
Plasma $\beta$ & $\frac{2.4\times10^{-12}n_{e}(T_{e}+T_{i}/Z)}{B^{2}/(8\pi)}$ & $75$ & $3.8\times10^{2}$\tabularnewline
\hline 
Kinetic energy/Thermal energy & $\frac{\frac{1}{2}\frac{m_{p}n_{e}A}{Z}u^{2}}{2.4\times10^{-12}n_{e}(T_{e}+T_{i}/Z)}$ & $12$ & $18$\tabularnewline
\hline 
Reynolds number $Rm$ & $uL/\eta$$\bigg(\eta=3.2\times10^{5}\frac{Z\ln\Lambda}{T_{e}^{3/2}}\bigg)$ & $1.6\times10^{4}$ & $1.0\times10^{4}$\tabularnewline
\hline 
Magnetic Reynolds number $Re$ & $uL/\nu$$\bigg(\eta=1.9\times10^{19}\frac{T_{\mathrm{i}}^{5/2}}{A^{1/2}Z^{3}n_{e}\ln\Lambda}\bigg)$ & $1.1\times10^{4}$ & $3.3\times10^{4}$\tabularnewline
\hline 
Biermann number $Bi$ & $\frac{euBL}{ck_{B}T_{e}}$ & $71$ & $13$\tabularnewline
\hline 
Hall number $\Omega_{H}$ & $\frac{4\pi en_{e}uL}{Bc}$ & $1.3\times10^{3}$ & $1.4\times10^{3}$\tabularnewline
\hline 
\end{tabular}
\end{table*}

\begin{table*}
\caption{Radiation properties of the jet at TCC for $d=800\mathrm{\mu m}$
ring radius. The temperature and density are in Table \ref{tab:Comparison-of-plasma}.\label{tab:Radiation-properties-of}}

\begin{tabular}{|c|c|c|}
\hline 
Plasma property & Formula & Value \tabularnewline
\hline 
\hline 
Planck opacity $\kappa_{P}$($\mathrm{cm^{2}/g}$) for CH target & from PROPACEOS & $1.8\times10^{-2}$\tabularnewline
\hline 
Optical depth $\tau$ for CH target & $\kappa_{P}\rho L$ & $1.1\times10^{-6}$\tabularnewline
\hline 
Cooling rate (1/s) for CH target & $0.72AZ^{-1}\kappa_{P}T_{e}^{3}$ & $3.2\times10^{6}$\tabularnewline
\hline 
Planck opacity $\kappa_{P}$($\mathrm{cm^{2}/g}$) for 2\% Fe dopant
target & from PROPACEOS & $3.9\times10^{-1}$\tabularnewline
\hline 
Optical depth $\tau$ for 2\% Fe dopant target & $\kappa_{P}\rho L$ & $4.4\times10^{-6}$\tabularnewline
\hline 
Cooling rate (1/s) for 2\% Fe dopant target & $0.72AZ^{-1}\kappa_{P}T_{e}^{3}$ & $3.2\times10^{7}$\tabularnewline
\hline 
\end{tabular}
\end{table*}

A list of on-axis plasma properties from a snapshot in FLASH simulation
results is listed in Table \ref{tab:Simulated-plasma-properties}
using the snapshot for $d=800\mu$m case at $t=$3ns. Other relevant
physical terms can be deduced from the scales and dimensionless numbers
in Table \ref{tab:Simulated-plasma-properties}. The plasma in the
jet is fully ionized, i.e $A=6.5$ and $Z=3.5$ for non-doped shots,
$A=7.49$ and $Z=3.95$ for doped shots. The optical Thomson scattering
diagnostics are simulated and discussed in Sec. \ref{subsec:comp_ts}.
By including laser energy deposition from the probe beam, the hydrodynamical
variables in a small region of around $(100\mathrm{\mu m})^{3}$ will
change significantly. This effect is significant for the analysis
of diagnostics, but not of the main interest in the dynamical evolution
of the jet.

\subsection{Magnetic fields\label{subsec:simulation_results_Magnetic-fields}}

Without any initial magnetic fields, the seed magnetic filed is generated
via the Biermann battery term caused by the individual beam heating.
The azimuthal asymmetry in the system is significant for the generation
of seed fields. In table \ref{tab:Simulated-plasma-properties}, the
Hall number $\Omega_{H}$ is much larger than the Biermann number
$Bi$. The Hall term is zero if $\boldsymbol{B}=0$, so it does not
generate seed fields. Thus the Hall term ($-\frac{c}{e}\nabla\times\frac{(\nabla\boldsymbol{B})\times\boldsymbol{B}}{4\pi n_{e}}$)
is neglectable in our system. Biermann battery term is the only source
term in the generalized Ohm's law that we calculate in FLASH simulation.
Magnetic resistivity is also included in the computation. However,
due to large magnetic Reynolds number, magnetic reconnection can hardly
happen until later on in the MHD picture. 

\begin{figure*}
\subfloat[]{\includegraphics[scale=0.15]{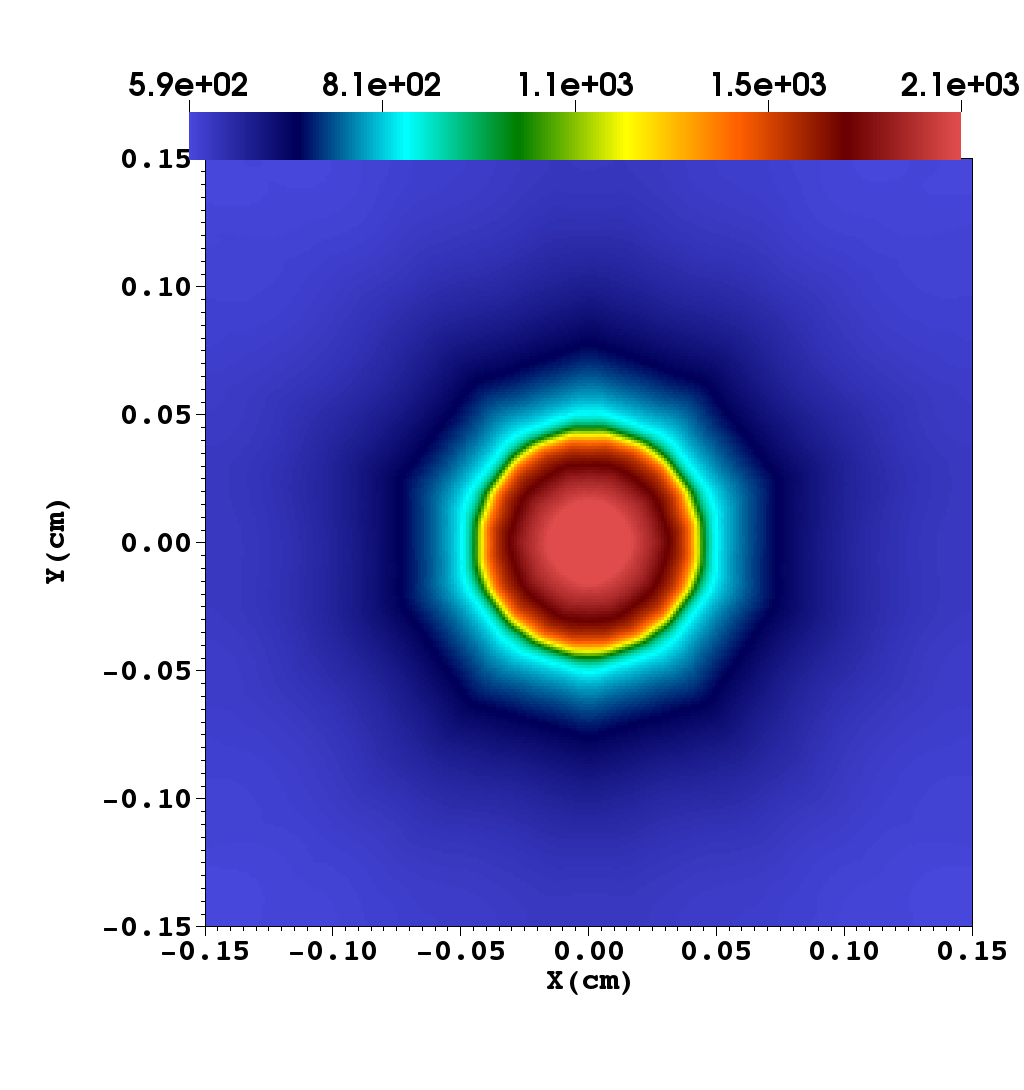}

}\subfloat[]{\includegraphics[scale=0.15]{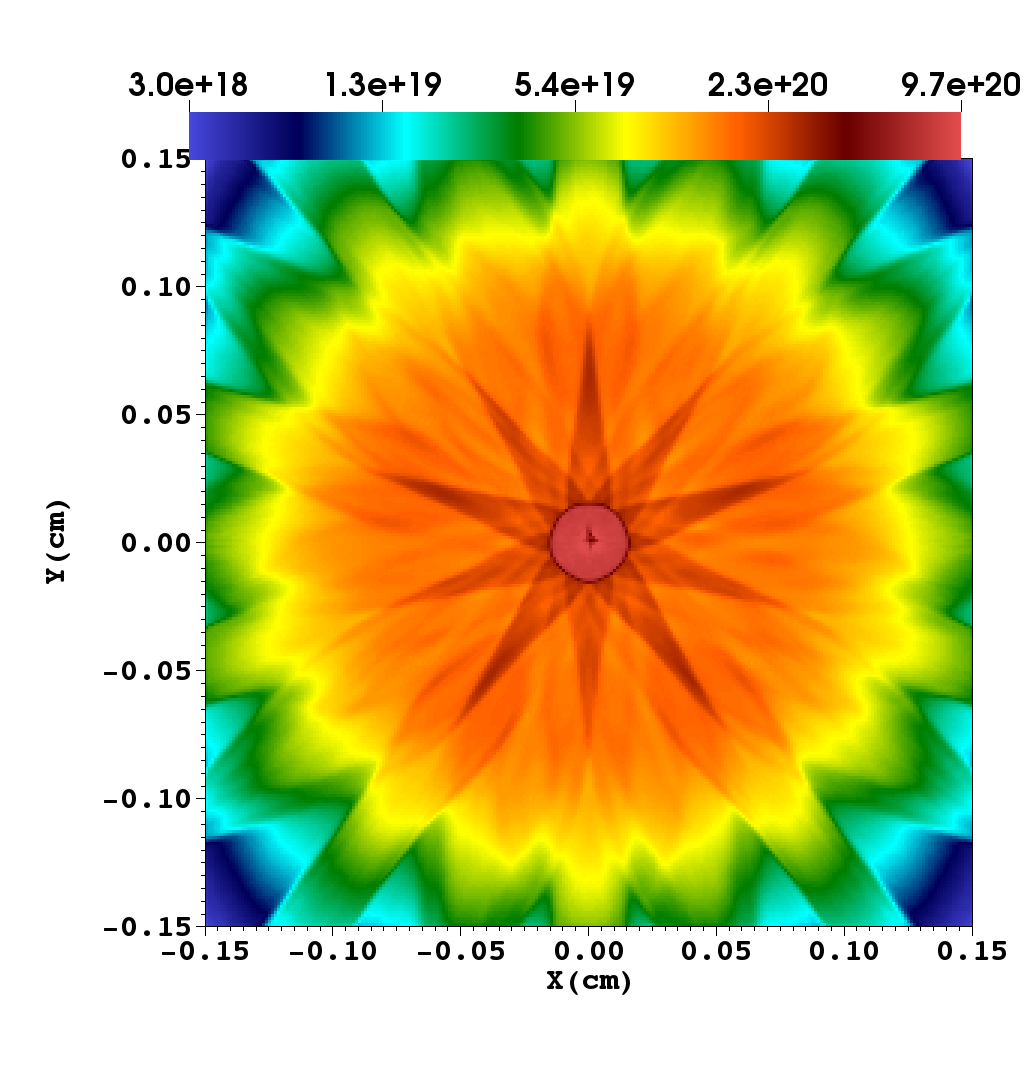}

}\subfloat[]{\includegraphics[scale=0.15]{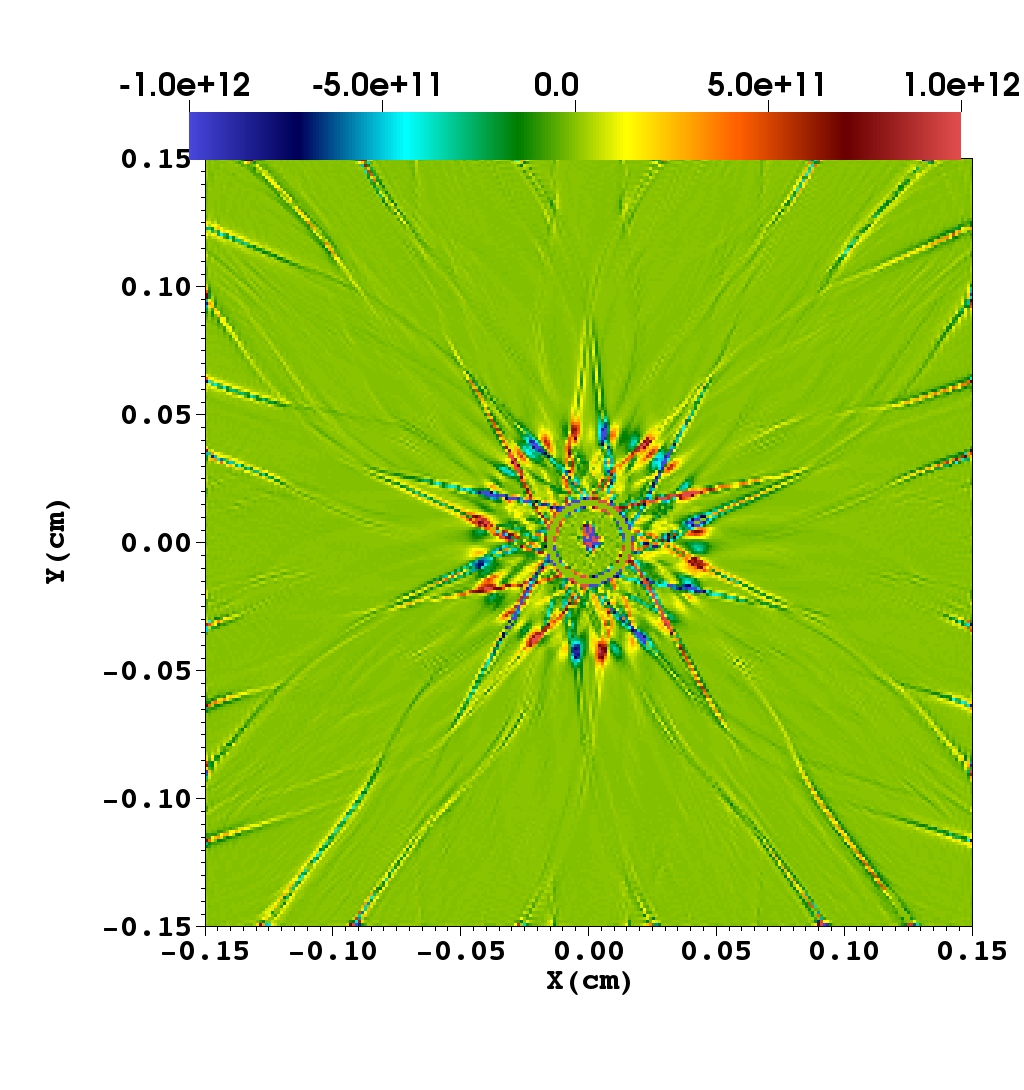}

}

\subfloat[]{\includegraphics[scale=0.15]{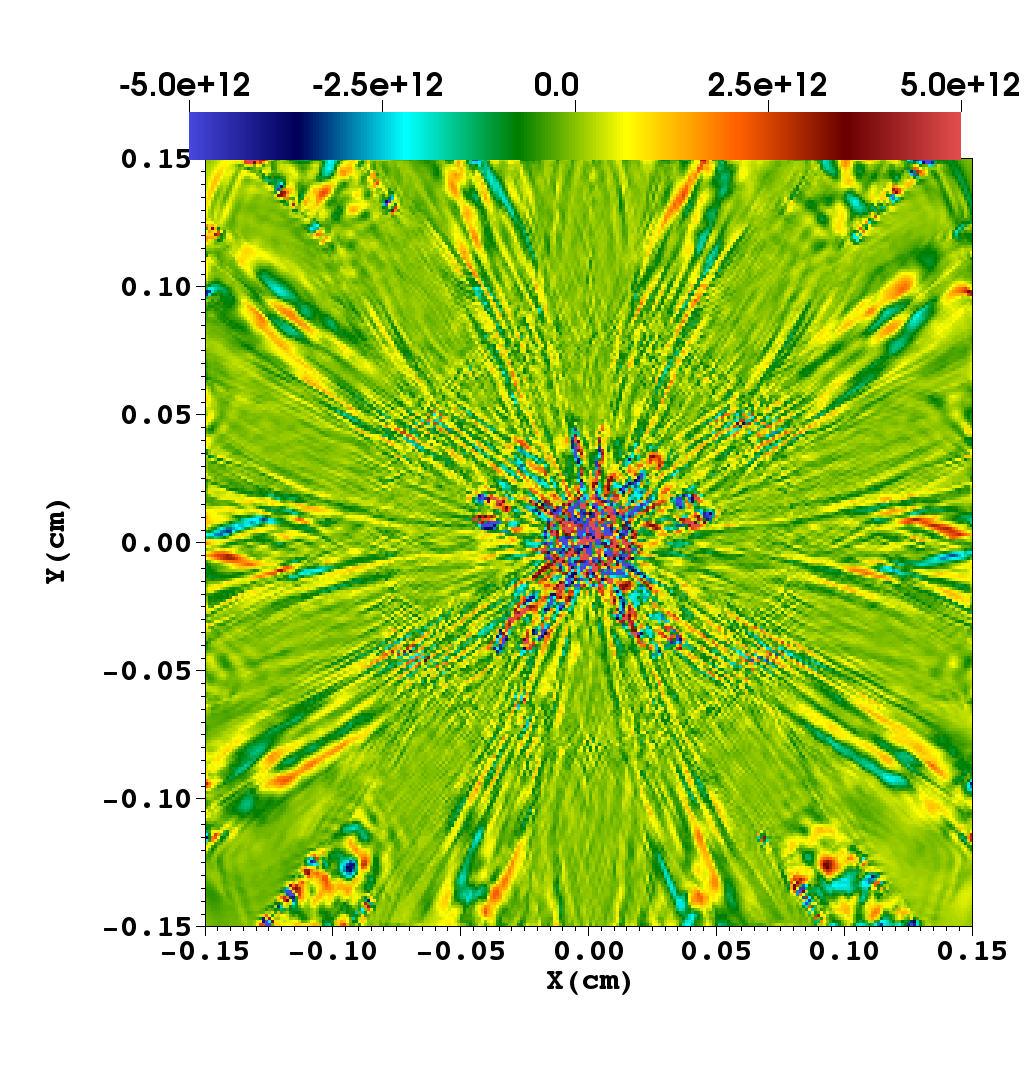}

}\subfloat[]{\includegraphics[scale=0.15]{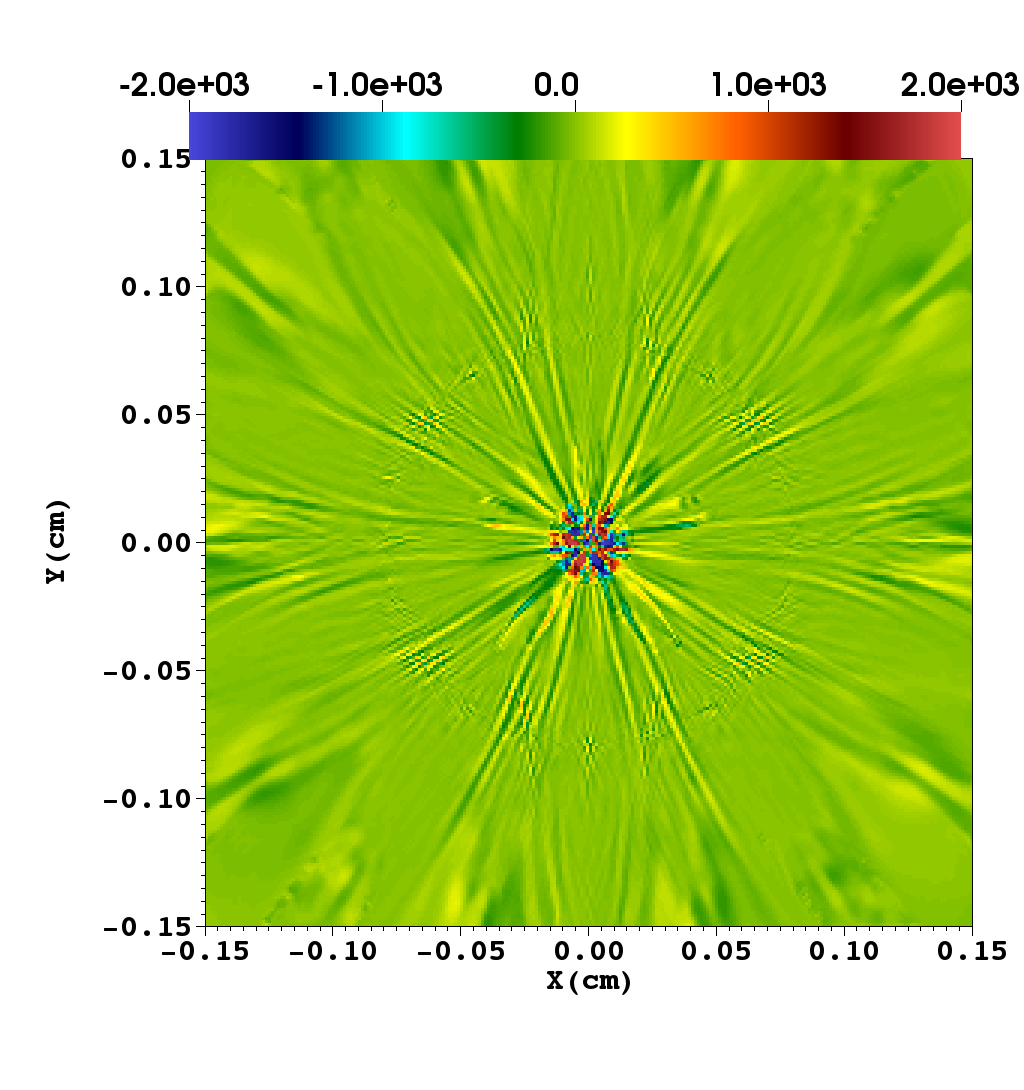}

}\subfloat[]{\includegraphics[scale=0.15]{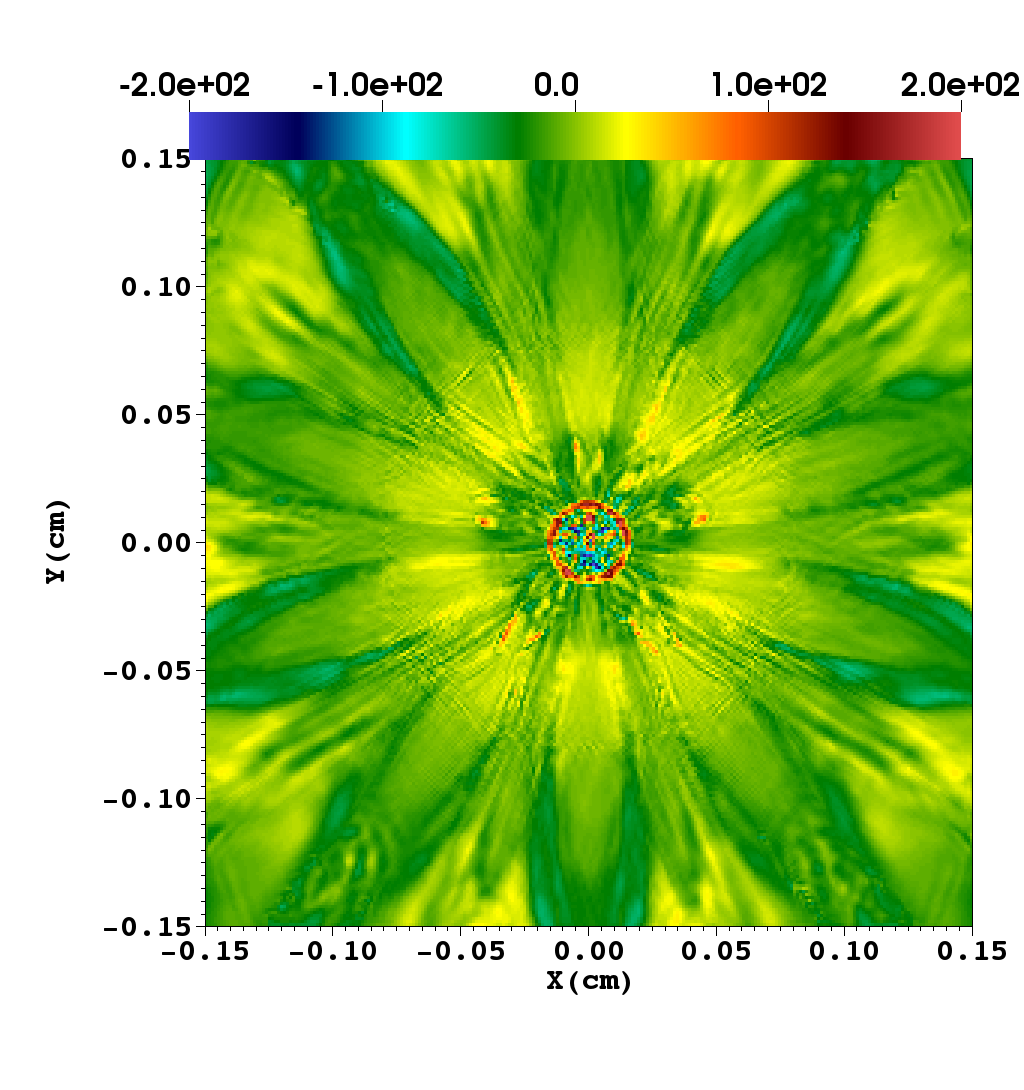}

}

\caption{Slice plot of several quantities at $z=0.1\mathrm{cm}$ for $t=1.6\mathrm{ns}$.
These figures demonstrate the generation and evolution of the axial
dominant magnetic field with alternating polarity and 5-fold symmetry
(a) Electron temperature(eV). The pattern is concentric circles. (b)
Electron density($\mathrm{cm^{-3}}$). The ``sunflower-like'' pattern
has 5-fold symmetry due to the laser pattern. The symmetry is slightly
broken due to finite number of cells in the simulation (c) z component
of Biermann battery term $\frac{c}{e}\nabla\times\frac{\nabla P_{e}}{n_{e}}=\frac{ck_{B}}{e}\nabla T_{e}\times\nabla n_{e}$
(kG/s) (d) z component of advection term $\nabla\times(\boldsymbol{v}\times\boldsymbol{B})$
(kG/s) (e) $z$ component of magnetic field(kG) (f) $\varphi$ component
of the magnetic field(kG).\label{fig:Slice-at-tcc-b}}
\end{figure*}

\begin{figure}
\hfill{}\subfloat[]{\includegraphics[scale=0.1]{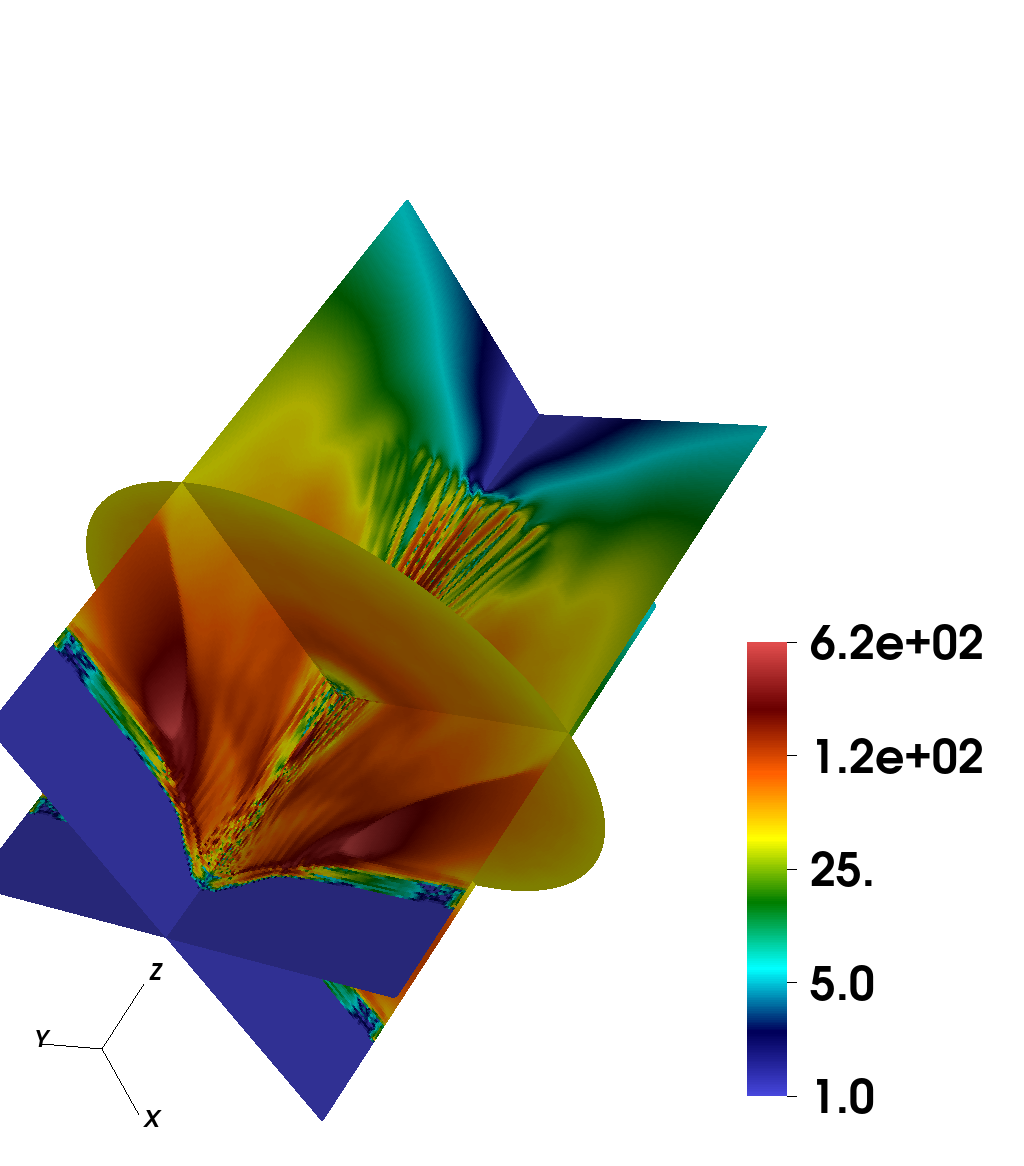}

}\hfill{}\subfloat[]{\includegraphics[scale=0.1]{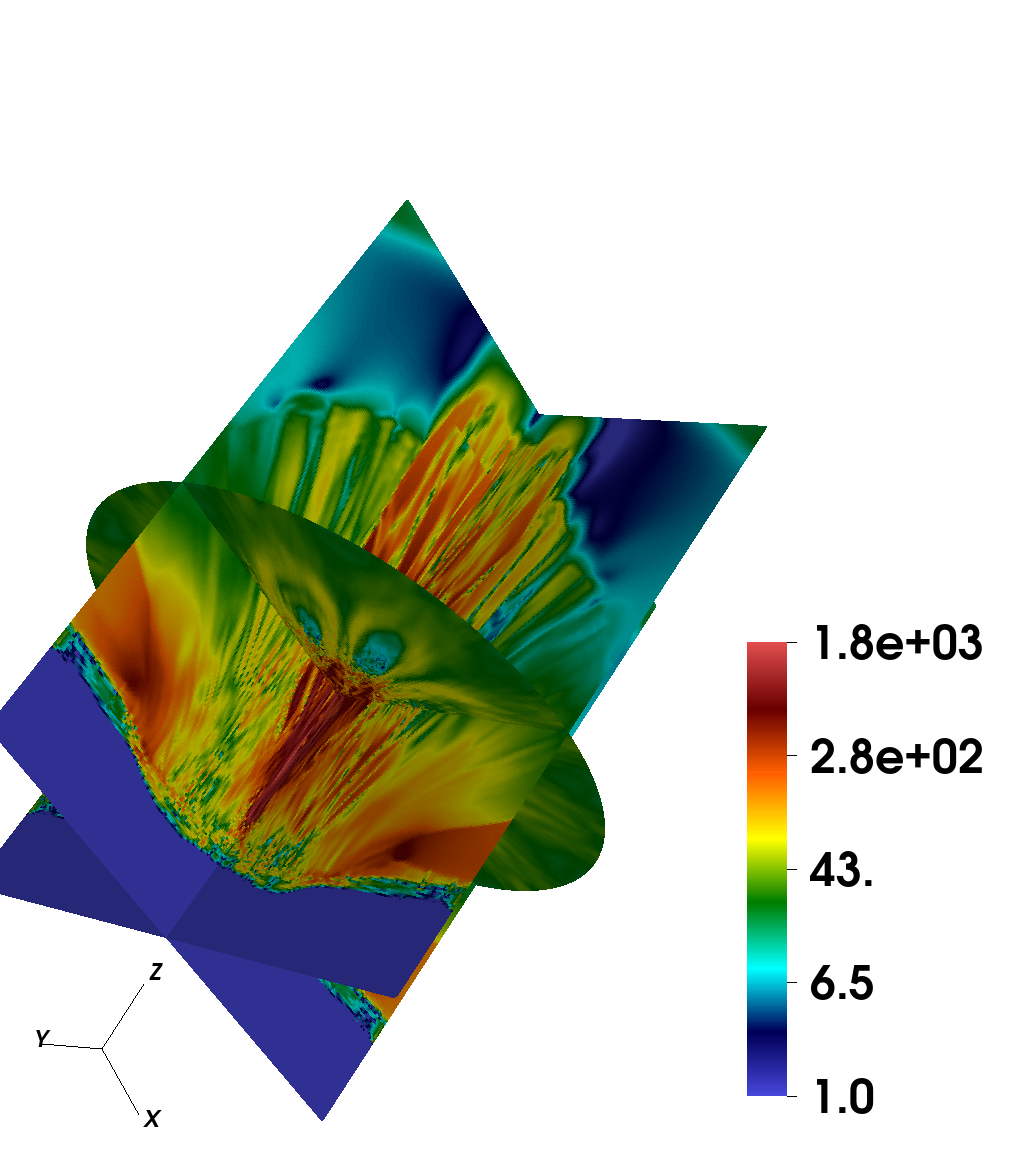}

}\hfill{}

\hfill{}\subfloat[]{\includegraphics[scale=0.1]{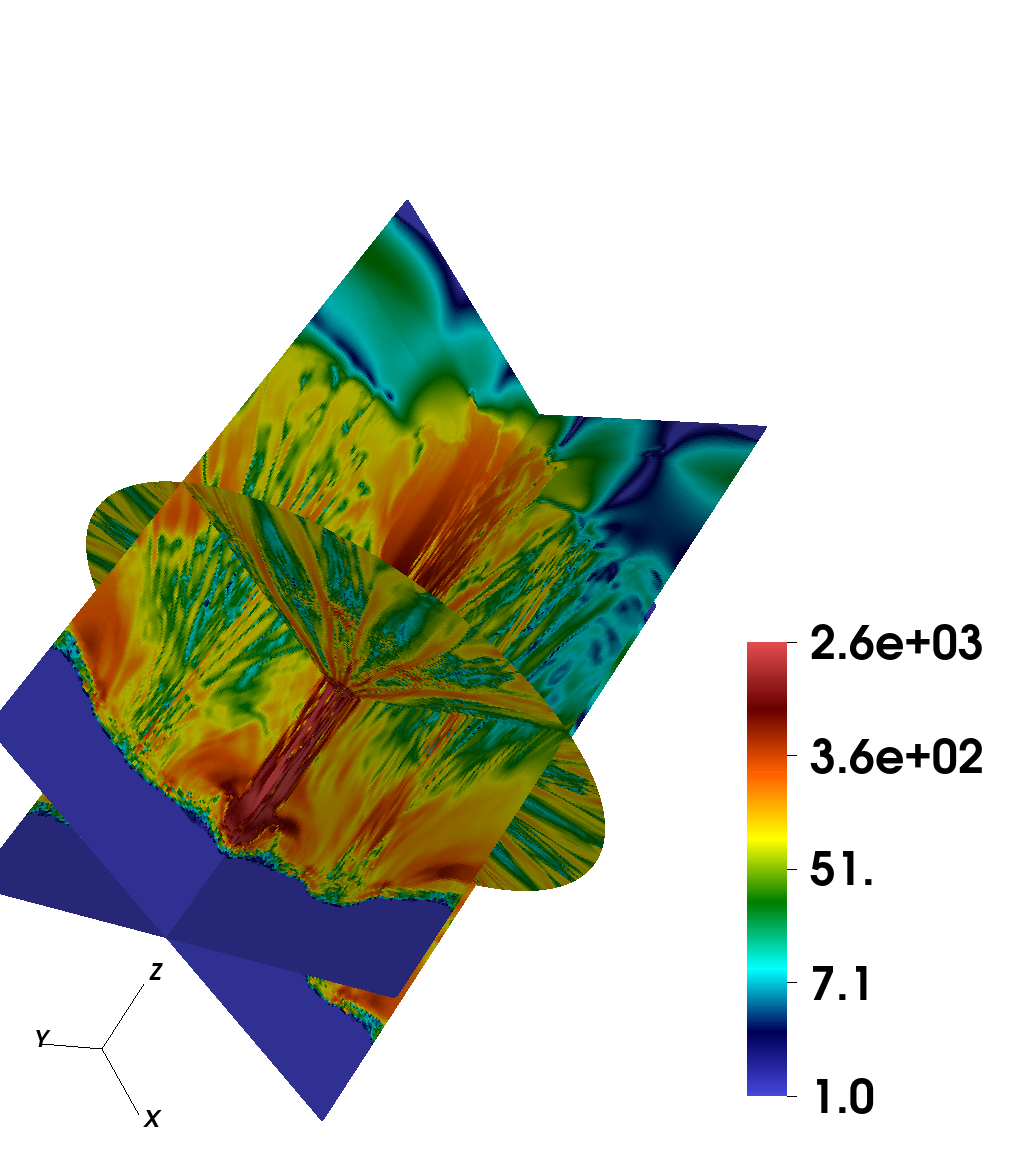}

}\hfill{}\subfloat[]{\includegraphics[scale=0.1]{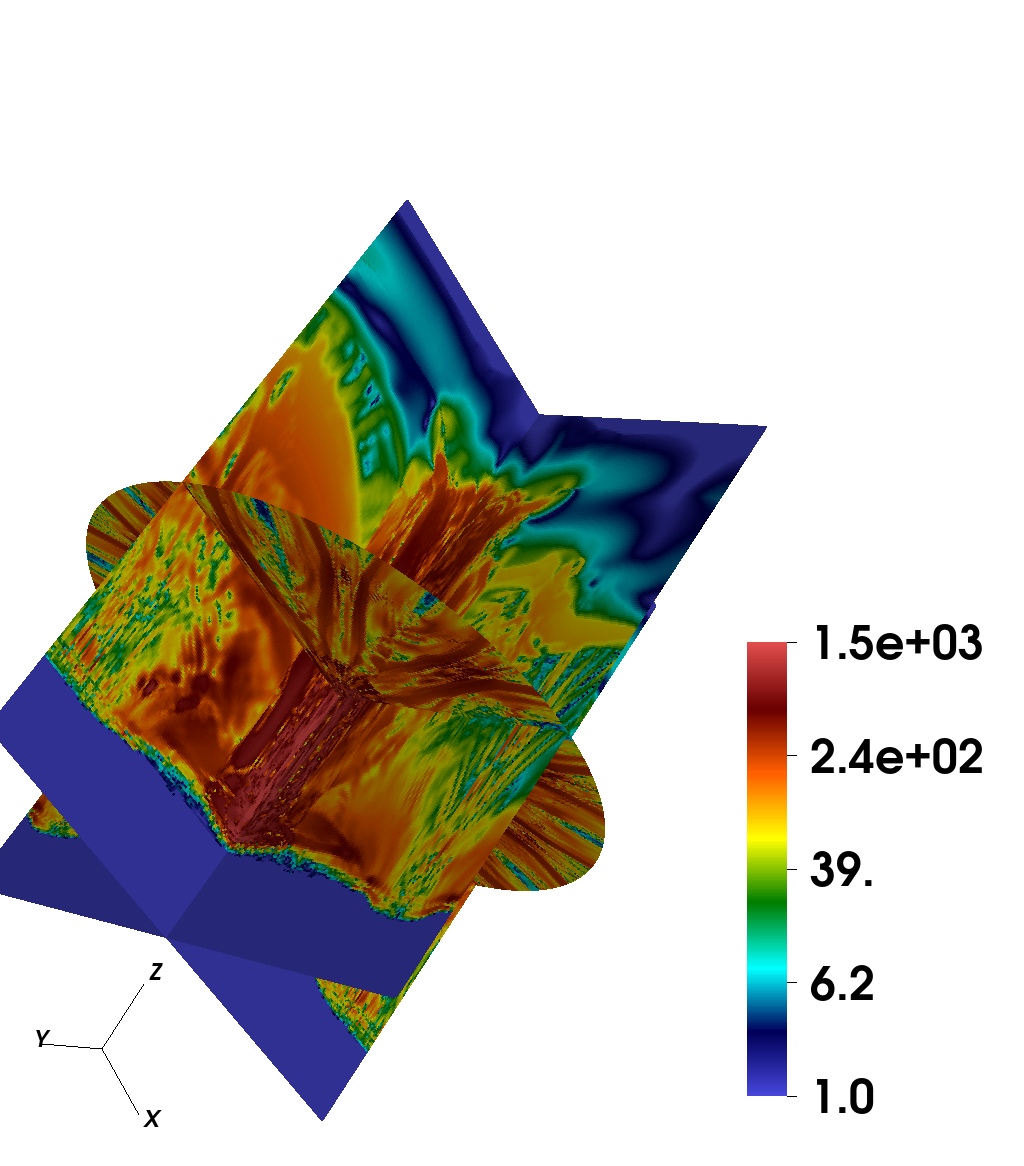}

}\hfill{}

\caption{Three slice plot for magnetic field amplitude(unit:kG) at $x=0$,
$y=0$ and $z=0.1\mathrm{cm}$ for different laser ring radius $d$
at $t=1.6\mathrm{ns}$. The disk slice is at $z=0.1\mathrm{cm}$ with
$0.3\mathrm{mm}$ diameter. (a) $d=0$ (b) $d=400\mathrm{\mu m}$
(c) $d=800\mathrm{\mu m}$ (d) $d=1200\mathrm{\mu m}$ \label{fig:mag_3slc_1.6ns}}
\end{figure}

\begin{figure}
\hfill{}\subfloat[]{\includegraphics[scale=0.1]{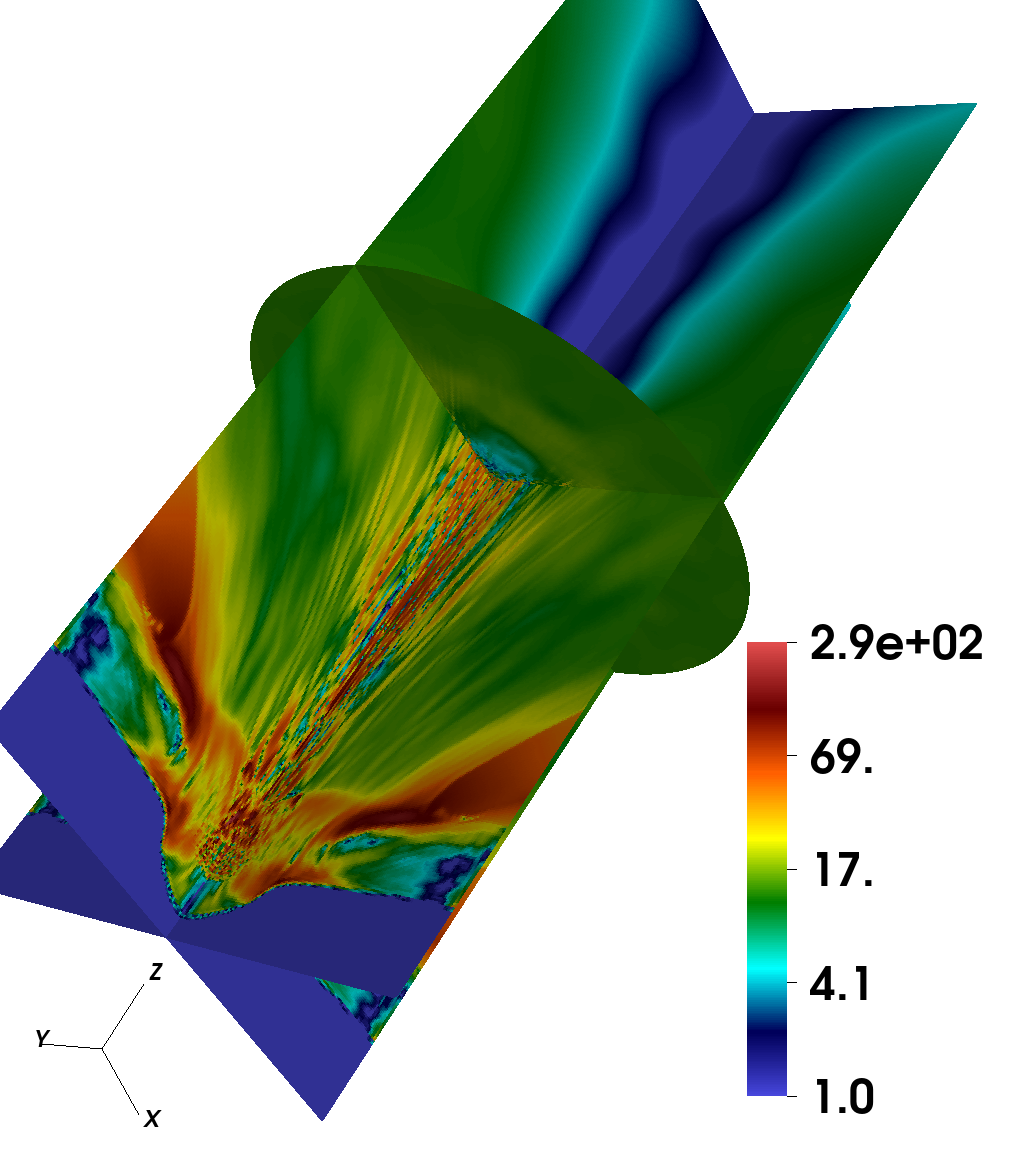}

}\hfill{}\subfloat[]{\includegraphics[scale=0.1]{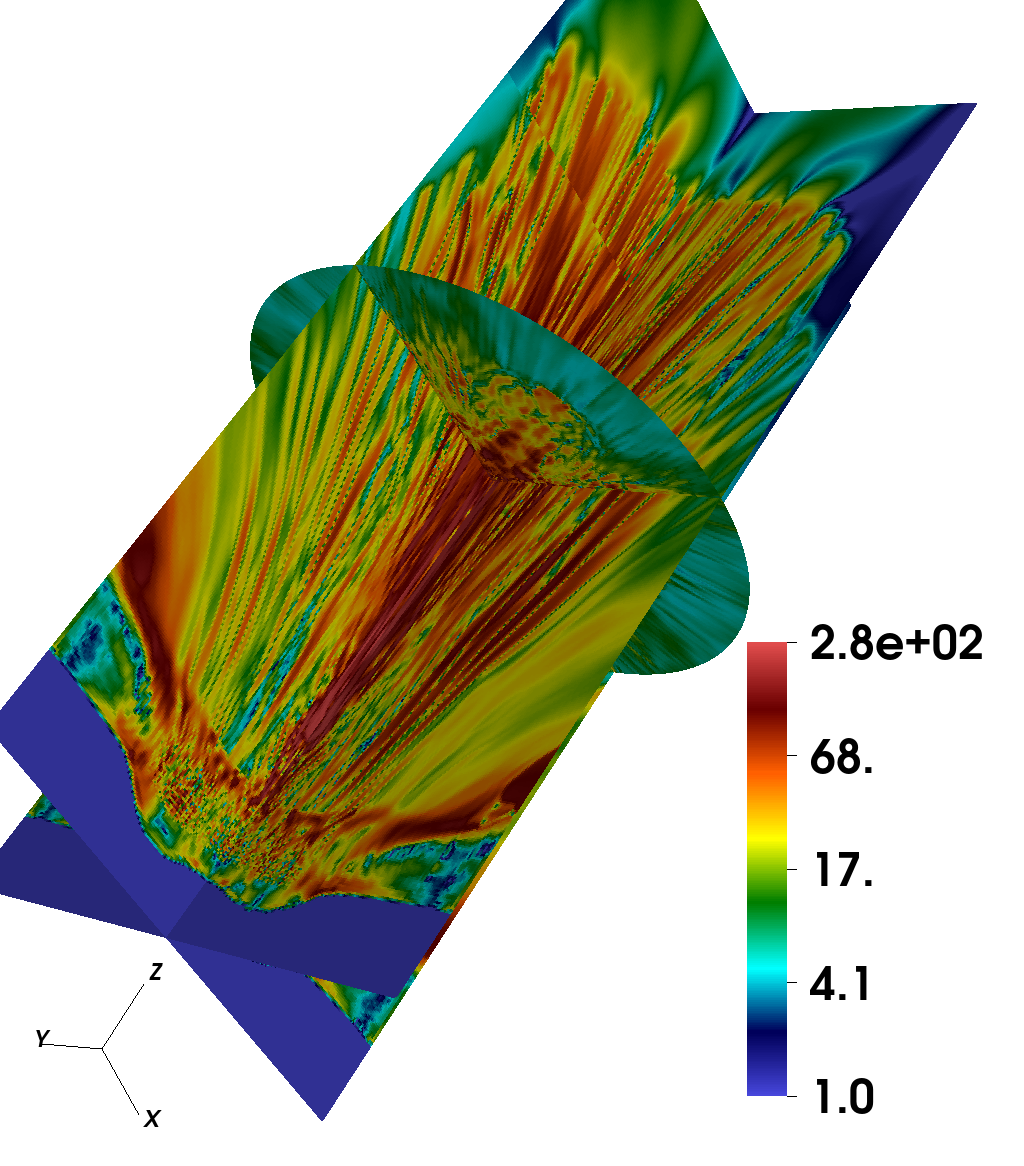}

}\hfill{}

\hfill{}\subfloat[]{\includegraphics[scale=0.1]{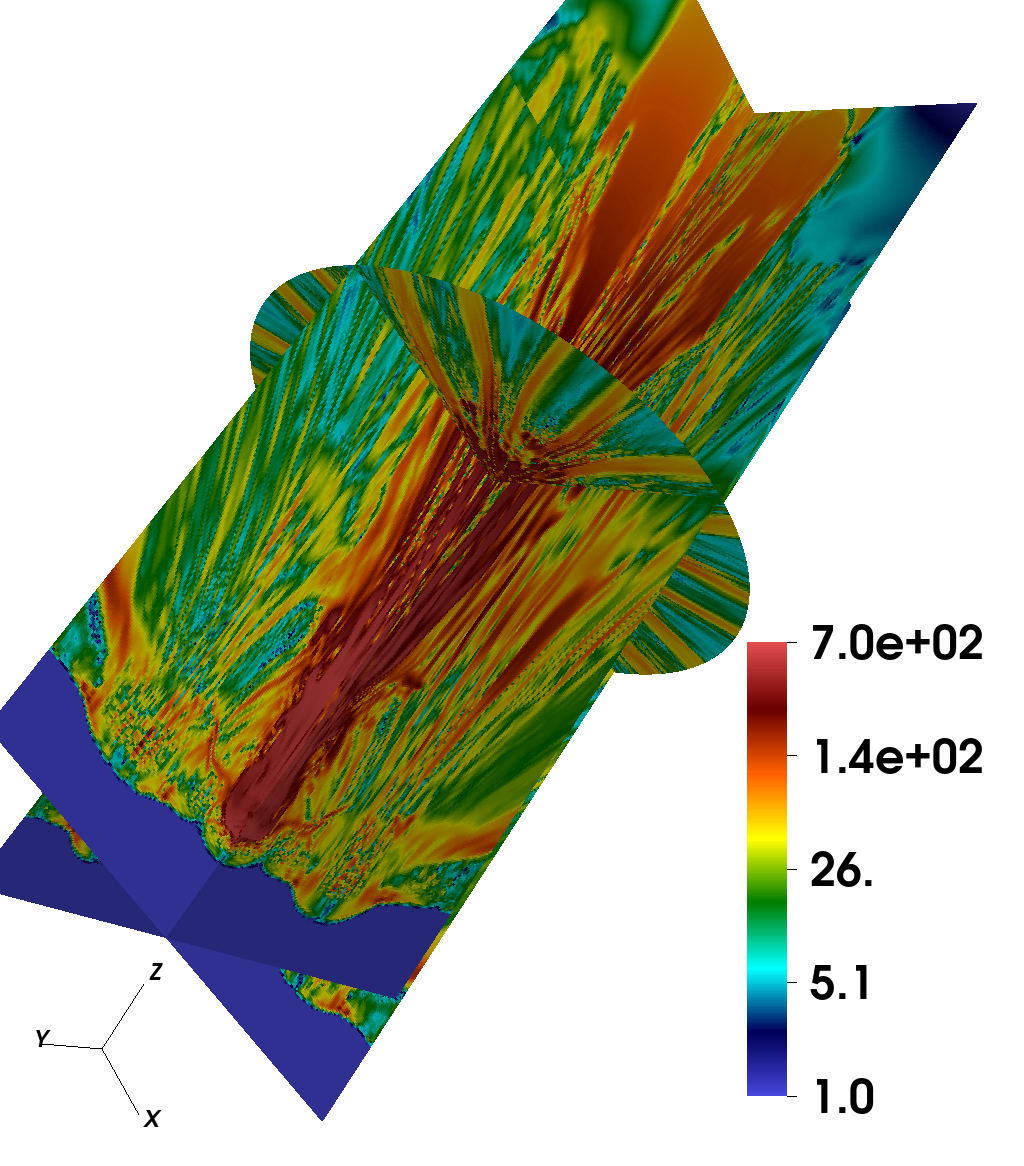}

}\hfill{}\subfloat[]{\includegraphics[scale=0.1]{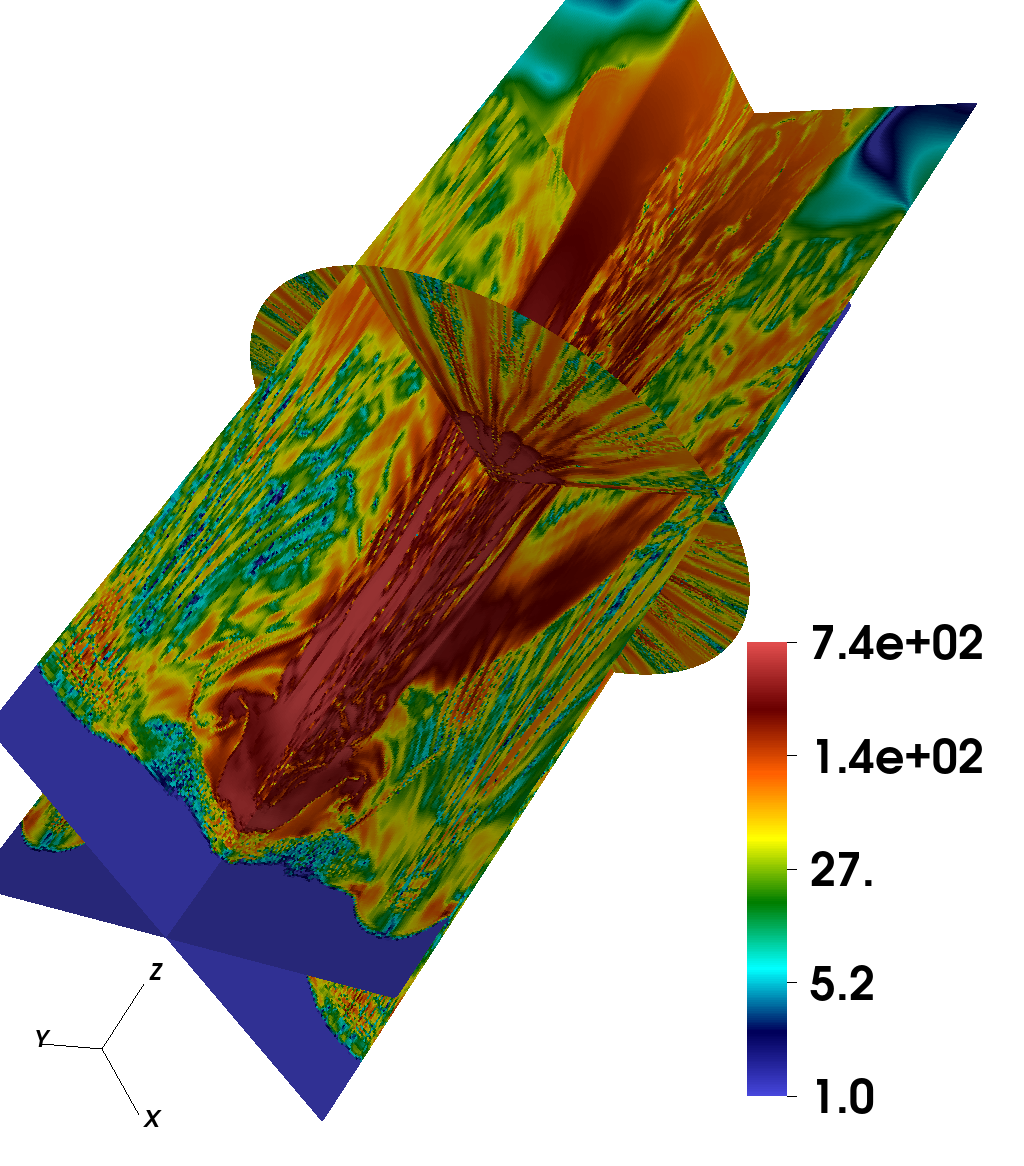}

}\hfill{}

\caption{Three slice plot for magnetic field amplitude(unit:kG) at $x=0$,
$y=0$ and $z=0.25\mathrm{cm}$ for different laser ring radius $d$
at $t=3.6\mathrm{ns}$. The disk slice is at $z=0.25\mathrm{cm}$
with $0.3\mathrm{mm}$ diameter. (a) $d=0$ (b) $d=400\mathrm{\mu m}$
(c) $d=800\mathrm{\mu m}$ (d) $d=1200\mathrm{\mu m}$.\label{fig:mag_3slc_3.6ns}}
\end{figure}

The generation and evolution of the axial dominant magnetic field
is demonstrated in Figure \ref{fig:Slice-at-tcc-b}. Because of the
radial temperature gradient(see Figure \ref{fig:Slice-at-tcc-b}(a))
and the azimuthal density gradient(see Figure \ref{fig:Slice-at-tcc-b}(b)),
the Biermann battery term ($\frac{c}{e}\nabla\times\frac{\nabla P_{e}}{n_{e}}=\frac{ck_{B}}{e}\nabla T_{e}\times\nabla n_{e}$)
is mainly in axial direction. Toroidal dominated magnetic fields are
only generated near the surface of the target, where there is little
azimuthal density gradient but large axial density gradient. At a
millimeter above the target surface, the magnetic field is generated
in the surrounding(the ring near $r\approx0.08\mathrm{cm}$ in Figure
\ref{fig:Slice-at-tcc-b}(c)) and advected into the core(the central
part of Figure \ref{fig:Slice-at-tcc-b}(d)). The shock amplifies
the axial magnetic field by a factor of $\sim4$ due to the flux conservation
as the plasma flows from the surrounding to the core. The cylindrical
shock makes the magnetic field highly concentrated as shown in Figure
\ref{fig:Slice-at-tcc-b}(e). Because the gradient of density alternates
several times azimuthally, the generated axial field also alternates.
The 5-fold symmetry in the field comes from the 5-fold symmetry in
arranging the laser spots as shown in Figure \ref{fig:The-illuminated-area}.
The symmetry is slightly broken in the simulation due to the cubic
cells and finite resolution. By using a larger ring of laser spots,
the magnetic energy is more concentrated in the core of the jet as
shown in Figure \ref{fig:mag_3slc_1.6ns} and \ref{fig:mag_3slc_3.6ns}. 

\begin{figure}
\subfloat[]{\includegraphics[scale=0.2]{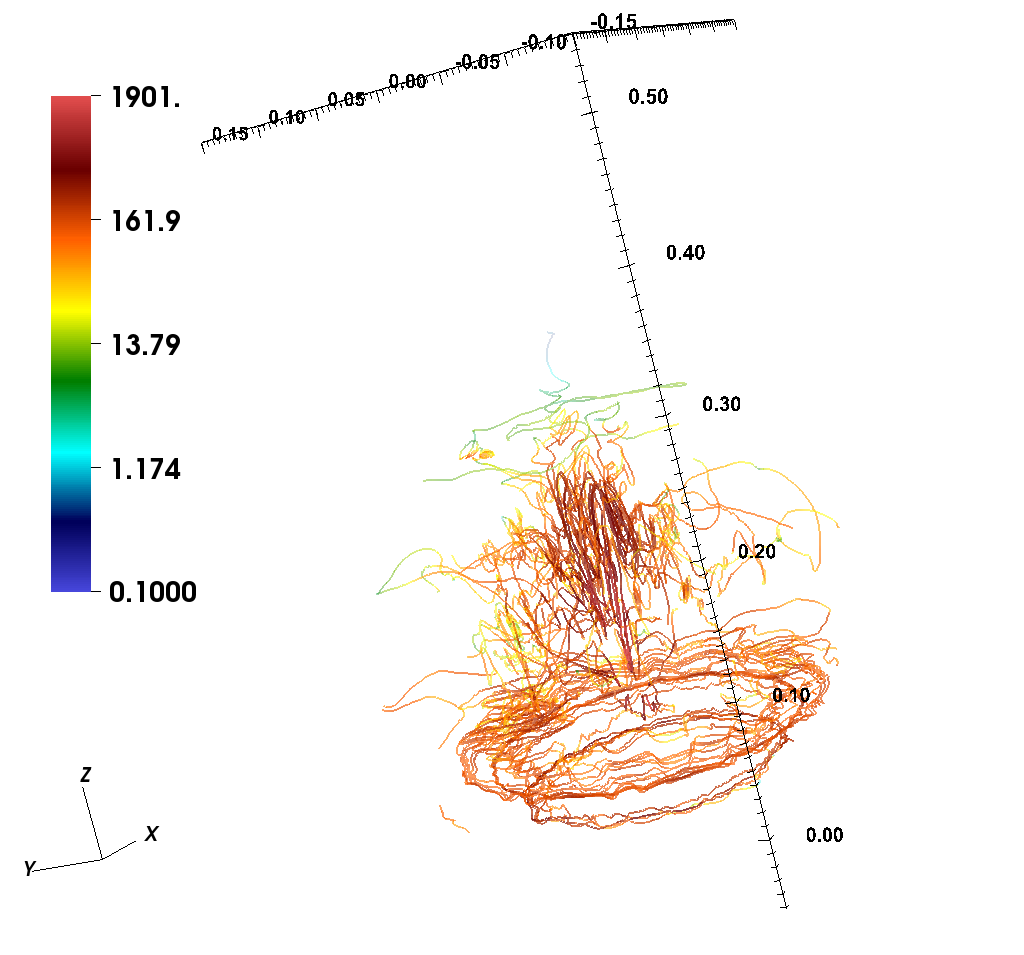}

}

\subfloat[]{\includegraphics[scale=0.2]{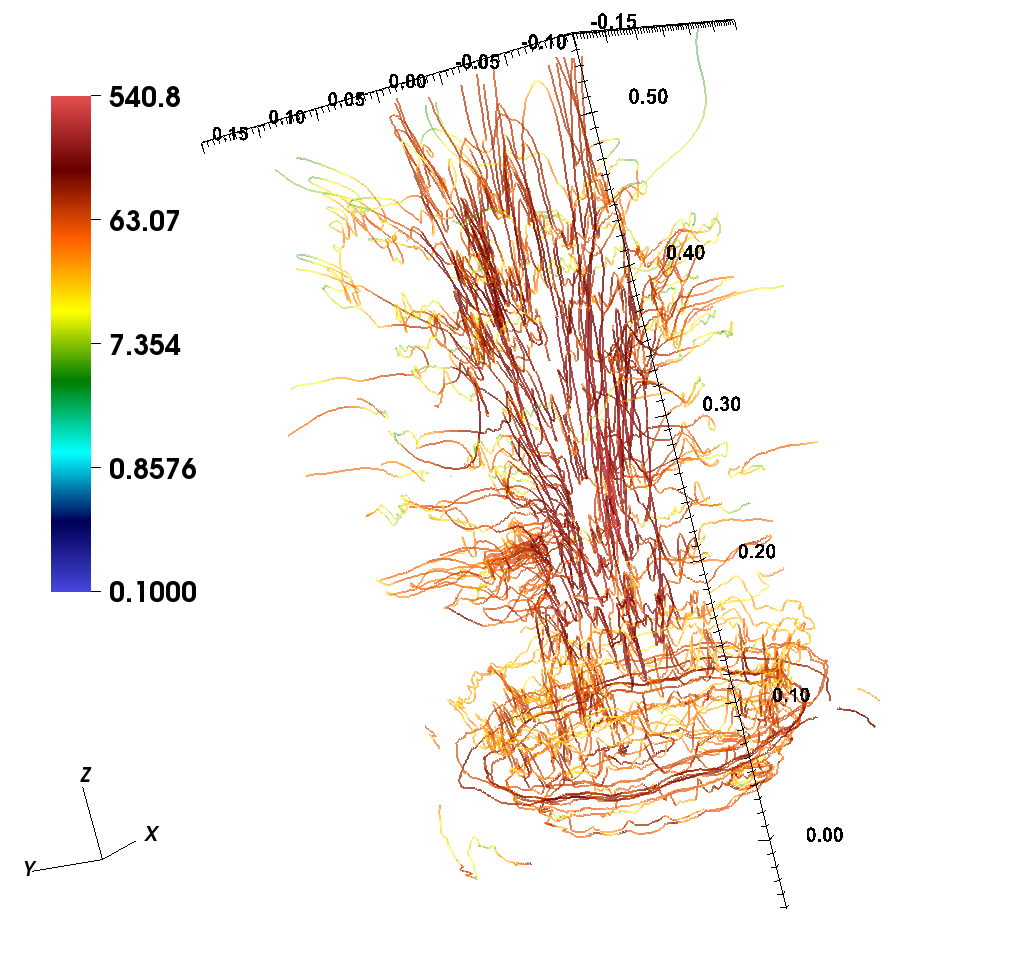}

}

\caption{Sample magnetic field lines(color scale unit: kG) for $d=800\mathrm{cm}$,
CH target. (a) at t=1.6ns (b) at t=3.6ns. The field far way from the
target is mainly axial and the field close the the target is toroidal.\label{fig:Magnetic-field-lines}}
\end{figure}

The magnetic field is mostly axial near the $z$ axis and mostly toroidal
near the surface of the target, as shown in Figure \ref{fig:Slice-at-tcc-b}(f)
and Figure \ref{fig:Magnetic-field-lines}. The width and the length
of the field bundles grow with the jet. The maximum field strength
reaches several hundred kilo-gauss. The maximum magnitude of magnetic
field at $t=\mathrm{3.6}\mathrm{ns}$ increases with the radius $d$
of the laser ring, as shown in Figure \ref{fig:mag_3slc_3.6ns}. This
is consistent with the 2D cylindrical simulation\citep{Fu_Fu2015}.
However, the full three dimensional simulation predicts a magnetic
field axial polarized and much stronger than those in the two dimensional
cylindrical simulation. In the 2D cylindrical simulation, the laser
intensity is azimuthally uniform, thus the Biermann battery term only
has the toroidal component.

\section{\label{sec:Comparison-to-experiments}Diagnostics modeling and Comparison
to experiments}

\subsection{Optical Thomson scattering spectrum \label{subsec:comp_ts}}

\begin{figure*}
\subfloat[]{\includegraphics[scale=0.5]{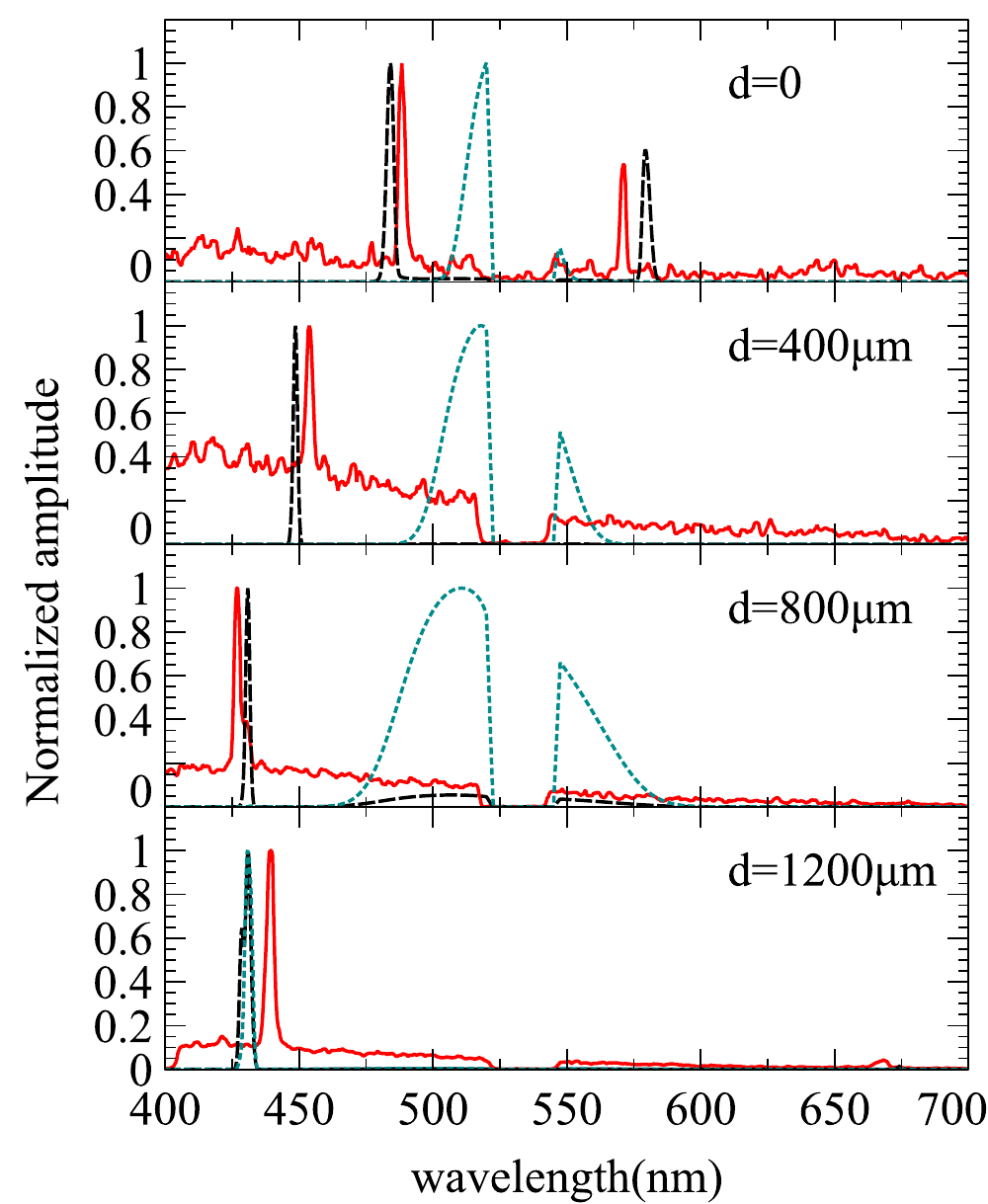}

}\includegraphics[scale=0.5]{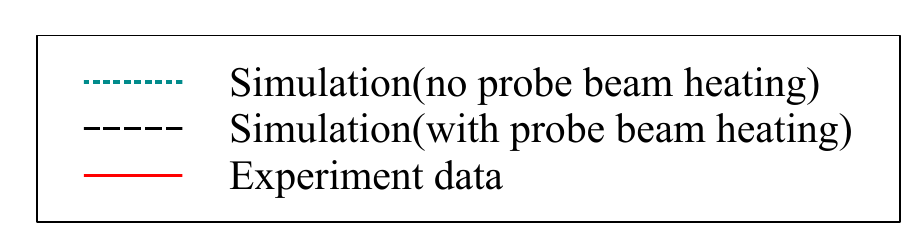}\subfloat[]{\includegraphics[scale=0.5]{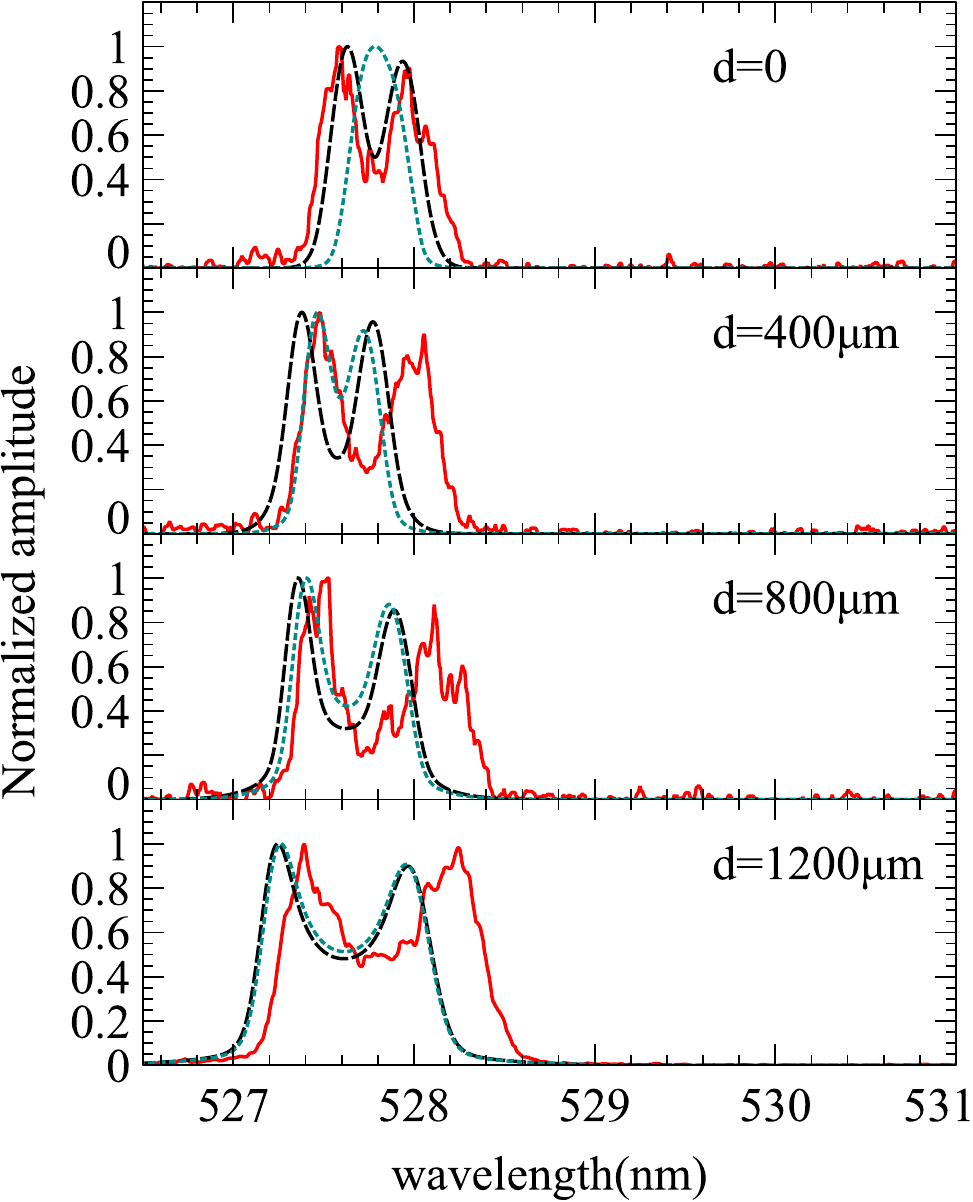}

}

\caption{Comparison between the synthetic optical Thomson-scattering spectra
based on FLASH simulations and the experimental data. The red solid
line is the experimental data, the blue dotted line is the synthetic
spectrum without probe beam heating, and the black dashed line is
the synthetic spectrum with probe beam heating. (a) EPW spectrum at
$3.6\mathrm{ns}$. (b) IAW spectrum at $3.9\mathrm{ns}$. \label{fig:ts_spect_compare}}
\end{figure*}

Although we can fit the optical Thomson-scattering spectrum using
the theoretical spectrum to infer the temperatures, density, and flow
velocity, the gradient of these quantities near TCC can affect the
spectrum and mislead the interpretation. As shown in Table \ref{tab:Simulated-plasma-properties},
the variation of some quantities can exceed $10\%$ and thus significantly
alters the spectrum. Moreover, the $2\omega$ probe beam can potentially
heat the plasma near TCC. In our simulation, the heating effect and
all the gradients are taken into account. Instead of directly comparing
the deduced quantities with those predicted in Figure \ref{fig:plasma_tcc},
we compare the synthetic spectrum with the data for experiment spectrum
in Figure \ref{fig:ts_spect_compare}. 

Figure \ref{fig:ts_spect_compare}(a) and (b) shows that the heating
from the TS probe has a significant impact on the measured spectra.
Although the energy in the probe beam ($25\sim50\mathrm{J}$) is low
compared to the drive beams, the $70\mathrm{\mu m}$ diameter focal
spot results in an intensity of $10^{15}\mathrm{W/cm^{2}}$. FLASH
simulations are performed with and without the probe beam to study
the impact of probe-beam heating. The locations of the TS peaks in
the simulated spectra that included the probe beam are in much better
agreement with the measured spectra. The effect is more pronounced
for smaller ring radii because the electron temperature is lower,
which leads to higher collisional absorption.

The background of the measured EPW spectrum comes from the bremsstrahlung
radiation, which is not calculated in the simulation. The bremsstrahlung
shape is apparent when the electron density is larger than $\sim10^{20}\mathrm{cm^{-3}}$.

\begin{figure}
\includegraphics[scale=0.5]{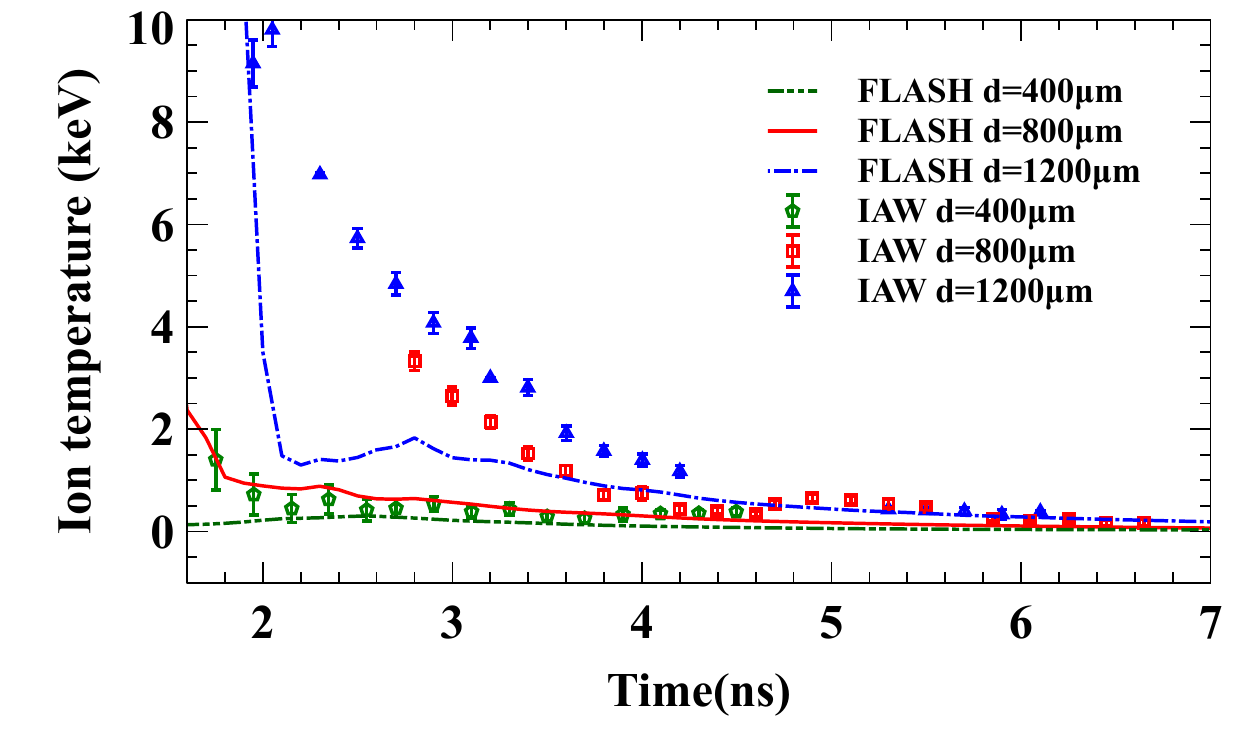}

\caption{The data for ion temperature from fitting the IAW spectrum and the
$(200\mathrm{\mu m})^{3}$ averaged value in FLASH simulations, both
at TCC .\label{fig:tion_correction}}
\end{figure}

The agreement for IAW spectrum is excellent for $d=0$ when the heating
is included, as shown in the first plot in \ref{fig:ts_spect_compare}(b).
However, for finite $d$, the simulation always underestimates the
width of the broadened line. The depth of the valley in the middle
of the shape is corrected by including the heating effect, which can
be explained by the increasing of the electron temperature from probe
heating. The under-predicted width of the IAW spectrum indicates the
under-predicted ion temperature. Because ions are not directly heated
by the probe beam, we also compare the ion temperature from fitting
the IAW spectrum and the $(200\mathrm{\mu m})^{3}$ averaged value
in FLASH simulations in Figure \ref{fig:tion_correction}. 

One may argue that the reason for underestimating the IAW line width
is the inaccuracy of the RAGE-like (it is so named because it is identical
to the method implemented in the radiation hydrodynamics code RAGE\citep{RAGE_Gittings2008})
energy apportion in our modeling. The RAGE-like approach apportions
the work term among the ions, electrons, and radiation field in proportion
to the partial pressures of these components. It is physically accurate
in smooth flow, but does not distribute internal energy correctly
among the ions, electrons, and radiation field at shocks. For the
finite $d$ case, strong and multiple shocks are presented. There
are the shocks between the plumes generated by neighboring beams and
the cylindrical shock surrounding the core. The core is usually a
secondary downstream. The ion heating exists at all shocks, but is
not calculated accurately using the RAGE-like approach. The electron
temperature should be significantly overestimated if the energy apportion
between electrons and ions is inaccurate. However, the comparison
between the measured IAW spectrum and the synthetic spectrum with
probe heating does not suggest any significant overestimation of electron
temperature. The reason for this might be the usage of electron heat
conduction in FLASH simulation, which mitigates the inaccuracy of
energy apportion. The extra part of the ion thermal energy measured
by IAW spectrum can only come from part of the kinetic energy in the
axial bulk motion of the flow, since multiple shocks already convert
the kinetic energy of radial and toroidal bulk motion into thermal
energy, and the magnetic energy is little comparing to the thermal
energy and the kinetic energy. For example, 10\% of the bulk kinetic
energy density at $t=3\mathrm{ns}$ and $z=2.5\mathrm{mm}$ corresponds
to $k_{B}T_{i}=0.1\times\frac{m_{i}u^{2}}{2}\approx2\mathrm{keV}$.
It is likely that the turbulence is developed from the flow velocity
difference between the plumes generated from different laser spots
due the laser intensity difference between these spots. The plasma
has high Reynolds number, as shown in Table \ref{tab:Simulated-plasma-properties}.
The kinetic energy in turbulent motion does not have to be dissipated
into heat to make the IAW spectrum broader, as long as a significant
amount of turbulent kinetic energy is cascaded down to a scale below
the resolution of Thomason scattering, i.e. $\sim100\mathrm{\mu m}$.
We will study the turbulence effect in a future work.

\subsection{Proton radiography\label{subsec:comp_prad}}

The proton images are smeared by a few factors (1) Spatial smearing:
the finite size of the proton source, which is $\sim$45$\mathrm{\mu m}$
for the fusion protons and $\sim5\mathrm{\mu m}$ for the TNSA protons;
(2) Temporal smearing: the pulse duration of the proton source, which
is $\sim150\mathrm{ps}$ for fusion protons and $\mathrm{1ps}$ for
TNSA protons. The pulse duration $\Delta t$ causes the smearing at
length scale $\Delta l\sim v\Delta t$, where $v$ is the characteristic
speed of the plasma. For fusion protons, $\Delta l_{z}\sim160\mathrm{\mu m}$,
$\Delta l_{x,y}\sim13\mathrm{\mu m}$, using the velocities in Table
\ref{tab:Simulated-plasma-properties}. For TNSA protons with $E=10\mathrm{MeV}$,
$\Delta l_{z}\sim1\mathrm{\mu m}$, $\Delta l_{x,y}\sim0.15\mathrm{\mu m}$;
(3) Spectrum smearing: the energy variation $\Delta E$ of the source
proton. Derived from Eq(16) in Graziani et. al.\citep{Prad_Graziani2017},
the variation of deflection angle cause by $\Delta E$ is $\frac{\Delta E}{2E}$
times the deflection angle. If there is only spectrum smearing, assuming
the proton is shifted by $200\mathrm{\mu m}$(which is typical) seen
in the TCC frame, it is expected that the 10.2MeV TNSA protons with
$\frac{\Delta E}{2E}=\frac{3.79\mathrm{MeV/2}}{2\times10.2\mathrm{MeV}}$($\Delta E$
is half of the effective temperature)resolve the magnetic field at
$\sim20\mathrm{\mu m}$, DD protons with $\frac{\Delta E}{2E}=\frac{0.32\mathrm{MeV}}{2\times3\mathrm{MeV}}$
resolve magnetic field at $\sim11\mathrm{\mu m}$, and $\mathrm{D^{3}He}$
protons with $\frac{\Delta E}{2E}=\frac{0.67\mathrm{MeV}}{2\times14.7\mathrm{MeV}}$
resolve magnetic field at $\sim5\mathrm{\mu m}$. The energy gain
or lost from the electric field is estimated to be less than $0.1\mathrm{MeV}$,
which is negligible compare to the $\Delta E$ of the beam itself.
Overall, our FLASH simulation is able to resolve a smaller spatial
scale than the experiment.

\begin{figure}
\subfloat[]{\includegraphics[scale=0.75]{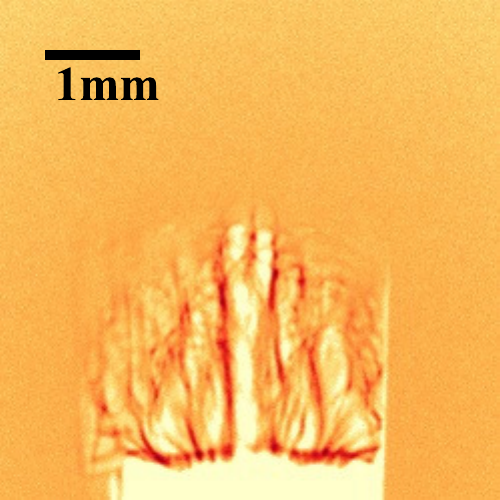}}\subfloat[]{\includegraphics[scale=0.75]{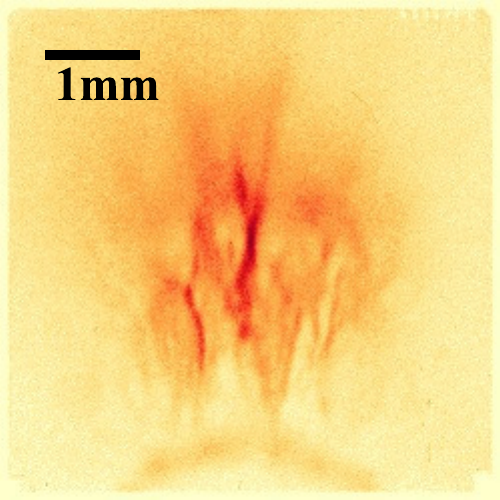}}

\subfloat[]{\includegraphics[scale=0.75]{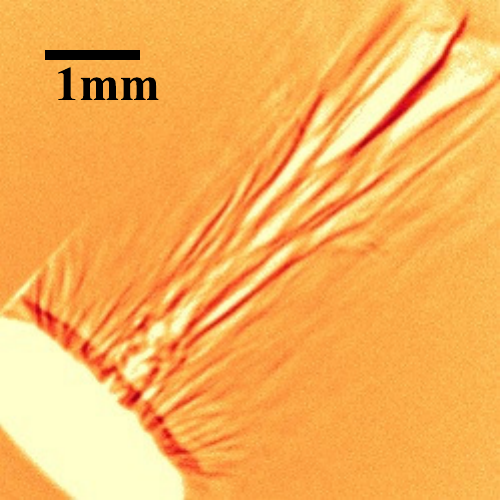}}\subfloat[]{\includegraphics[scale=0.75]{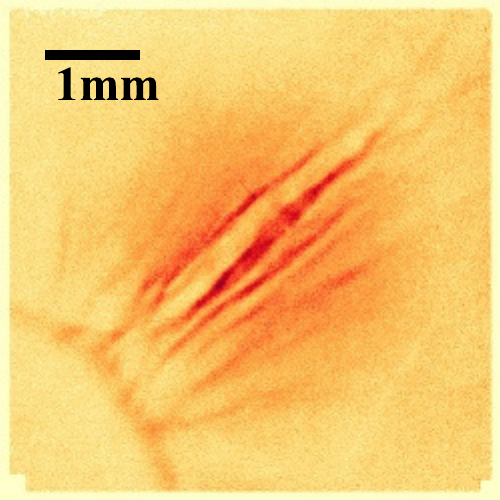}}

\caption{Comparison of synthetic proton image with data recorded on CR39 for
$14.7\mathrm{MeV}$ protons. The color scales are the same for all
images. The ring radius is $d=800\mathrm{\mu}\mathrm{m}$. The target
is CH without dopant. (a) synthetic image at $t=1.6\mathrm{ns}$,
the corresponding three-slice plot of the field is in Figure \ref{fig:mag_3slc_1.6ns}(c)
(b) experiment image at $t=1.6\mathrm{ns}$ (c) synthetic image at
$t=3.6\mathrm{ns}$, the corresponding three-slice plot of the field
is in Figure \ref{fig:mag_3slc_3.6ns}(c) (d) experiment image at
$t=3.6\mathrm{ns}$. The CR39 image plate is $\mathrm{10cm\times10cm}$.
On the plot, magnification is taken into account and the scale listed
is in the plasma frame. In (a) and (b), upward is the $+z$ direction.
In (c) and (d), upright is the $+z$ direction. The void region in
below in (a) and (b), and bottom left corner in (c) and (d) is the
target, which block the protons. \label{fig:D3He}}
\end{figure}

\begin{figure}
\subfloat[]{\includegraphics[scale=0.1]{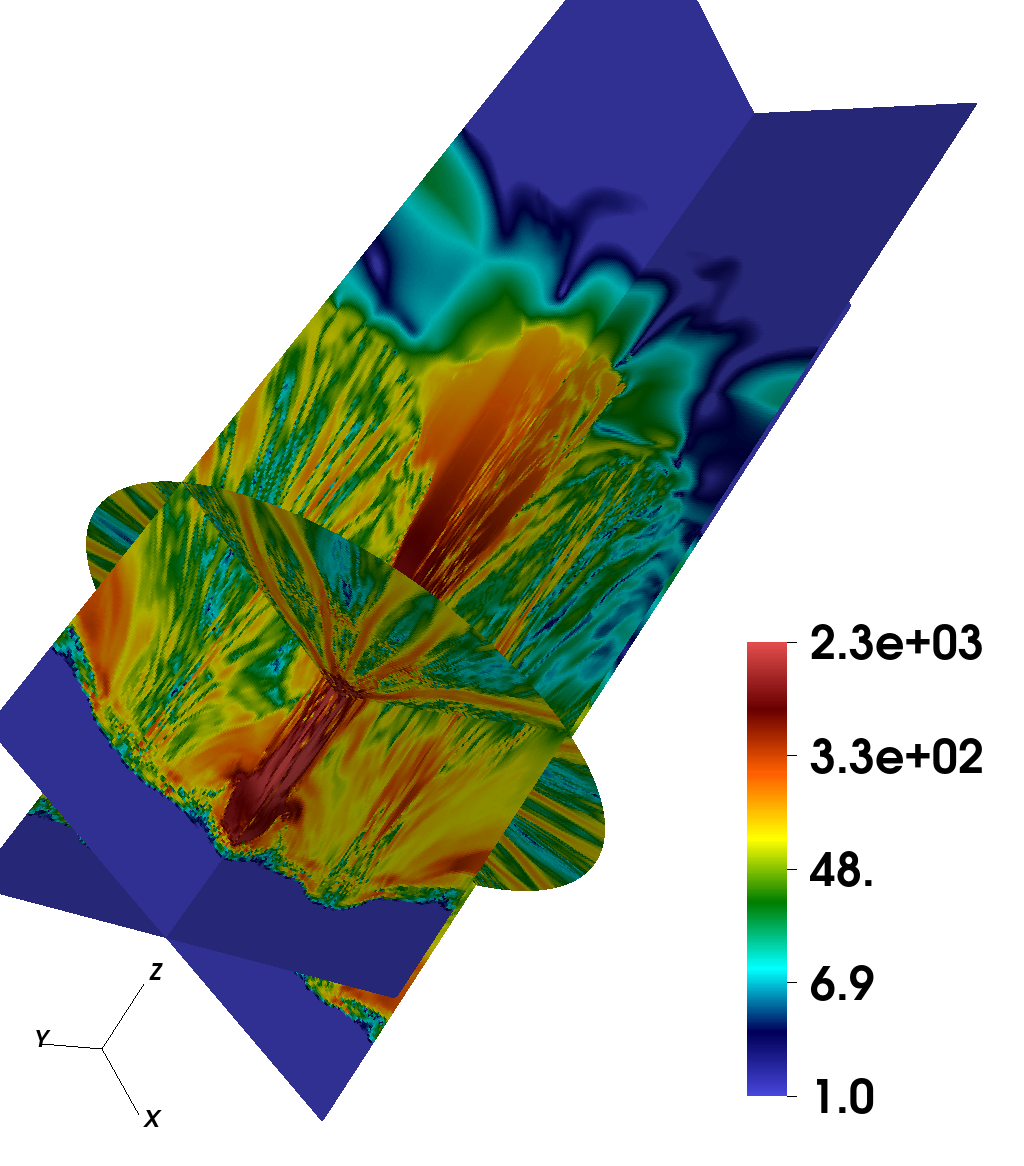}

}

\subfloat[]{\includegraphics[scale=0.75]{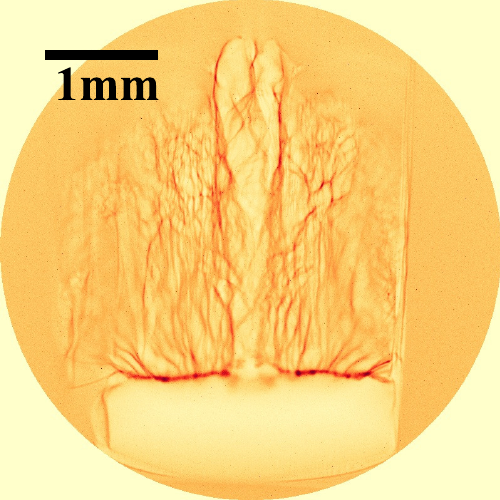}}\subfloat[]{\includegraphics[scale=0.75]{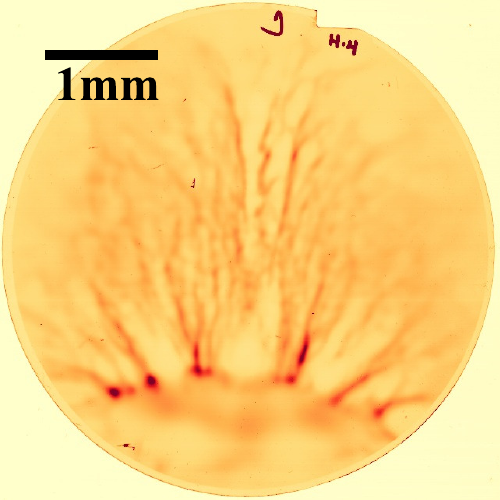}}

\caption{Comparison of synthetic proton image with data recorded on radiochromic
film for $10.2\mathrm{MeV}$ protons. The color scales are the same
for two images. The ring radius is $d=800\mathrm{\mu}\mathrm{m}$.
The target is CH without dopant. (a) three-slice plot at $x=0$, $y=0$,
$z=0.1\mathrm{cm}$, magnetic field strength(kG) plot at $t=\mathrm{1.9}\mathrm{ns}$
(b) synthetic image at $t=1.9\mathrm{ns}$ (c) experiment image at
$t=1.9\mathrm{ns}$ from H4 pack. The image plate is a disk of $10\mathrm{cm}$.
The scale and orientation is as same as Figure \ref{fig:D3He}(a)
and (b).\label{fig:EP_Proton}}
\end{figure}

\begin{figure*}
\hfill{}\hfill{}\subfloat[]{\includegraphics[scale=0.1]{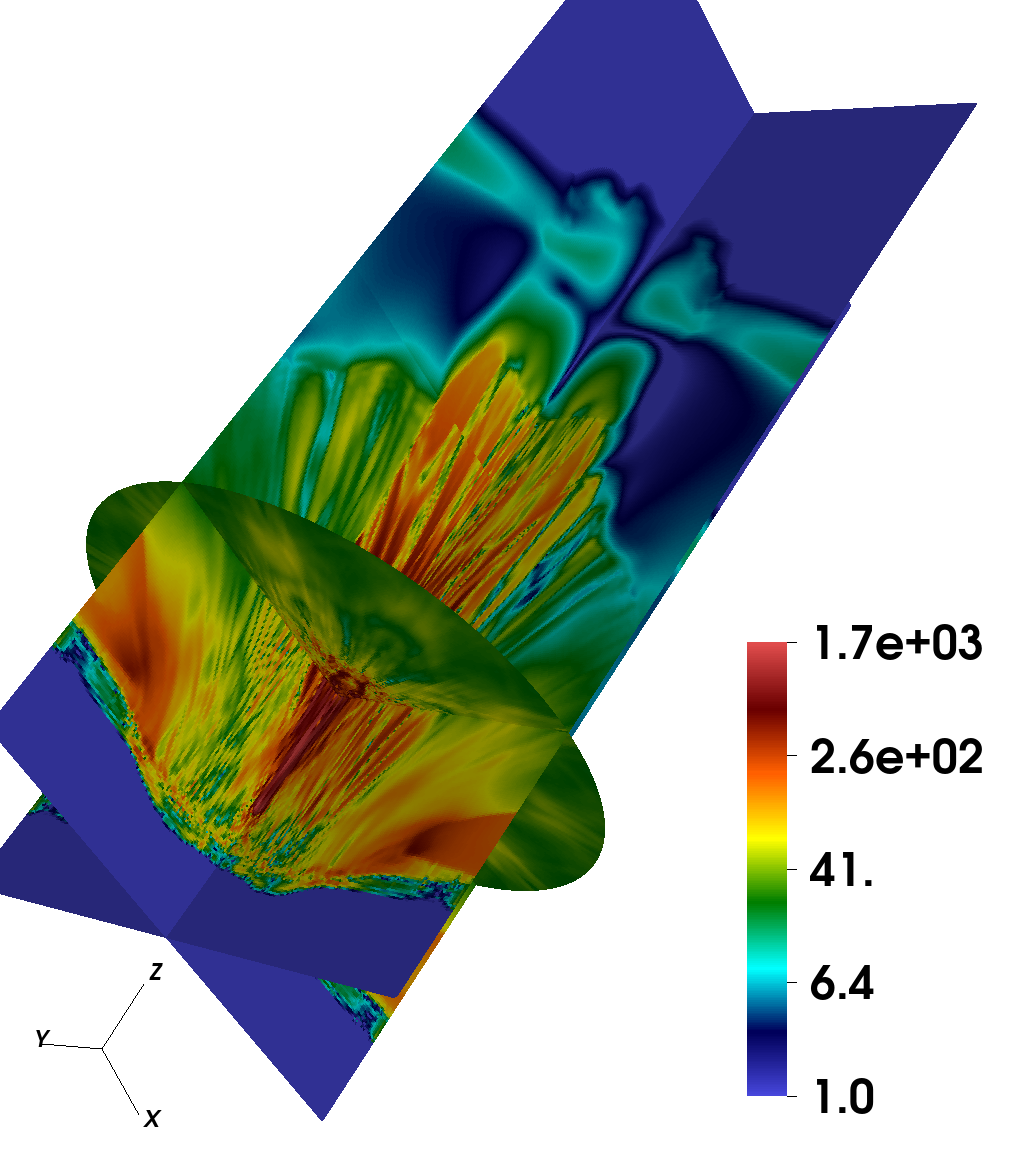}

}\hfill{}\subfloat[]{\includegraphics[scale=0.75]{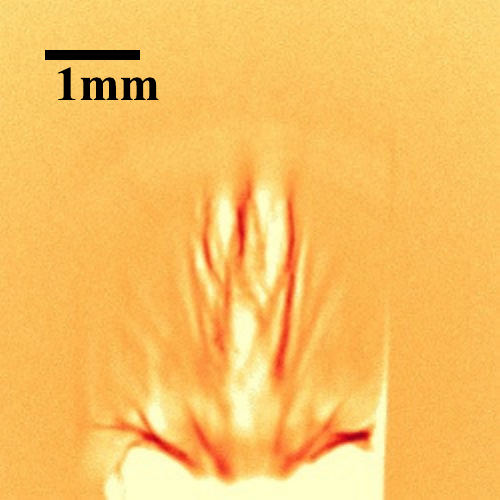}

}\hfill{}\subfloat[]{\includegraphics[scale=0.75]{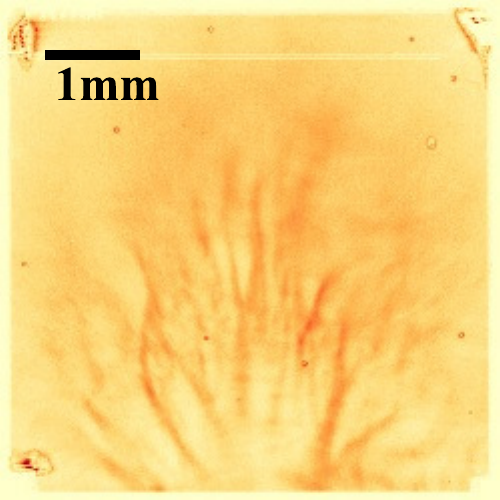}

}\hfill{}\hfill{}

\hfill{}\hfill{}\subfloat[]{\includegraphics[scale=0.1]{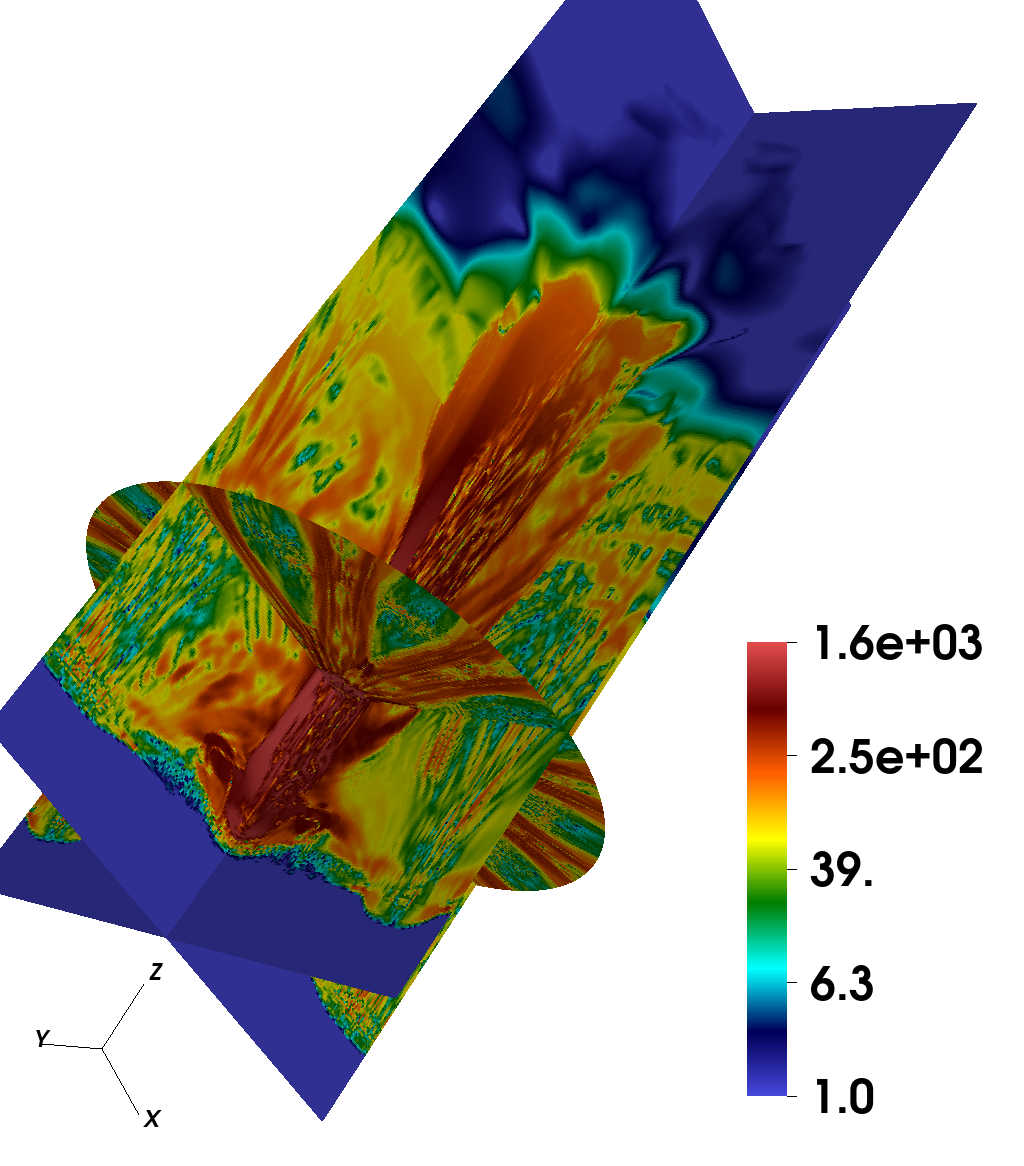}

}\hfill{}\subfloat[]{\includegraphics[scale=0.75]{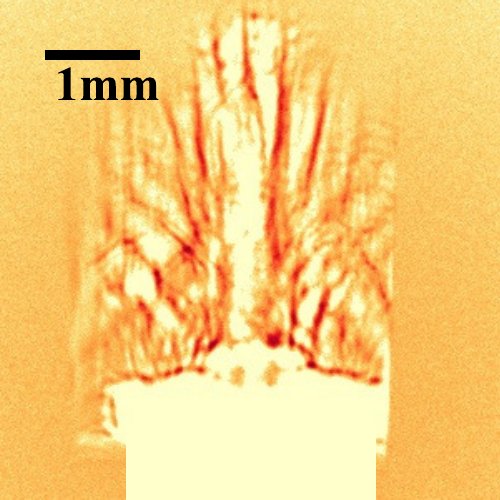}

}\hfill{}\subfloat[]{\includegraphics[scale=0.75]{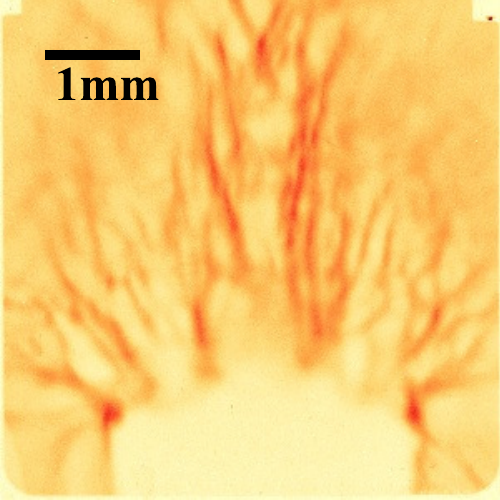}

}\hfill{}\hfill{}

\caption{(a) three-slice at $x=0$, $y=0$, $z=0.1\mathrm{cm}$, magnetic field
strength(kG) plot at $t=\mathrm{1.8}\mathrm{ns}$ for $d=400\mathrm{\mu m}$
(b) synthetic $3\mathrm{MeV}$ proton image at $t=1.8\mathrm{ns}$
for $d=400\mathrm{\mu m}$ (c) experiment $3\mathrm{MeV}$ proton
image image at $t=1.9\mathrm{ns}$ for $d=400\mathrm{\mu m}$ (c)
three-slice at $x=0$, $y=0$, $z=0.1\mathrm{cm}$, magnetic field
strength(kG) plot at $t=2.3\mathrm{ns}$ for $d=120\mathrm{\mu m}$
(d) synthetic $14.7\mathrm{MeV}$ proton image at $t=2.3\mathrm{ns}$
for $d=1200\mathrm{\mu m}$ (e) experiment $14.7\mathrm{MeV}$ proton
image image at $t=2.3\mathrm{ns}$ for $d=1200\mathrm{\mu m}$. The
scale and orientation of the image are as same as Figure \ref{fig:D3He}(a)
and (b). The color scales are the same for all images. \label{fig:prad_othercase}}
\end{figure*}

The simulation of images predicts the features observed in the experimental
data. To the lowest order the light and dark patterns correspond to
the averaged MHD current($\nabla\times\boldsymbol{B}$) projected
along the light of sight\citep{Prad_Graziani2017}. The alternating
axial field filaments result in several vertical dark and bright strips.
The curved horizontal strip close to the surface of the target is
produced by the large loop of surface toroidal field. Figure \ref{fig:D3He}
shows the comparison between the simulation synthetic and experimental
$\mathrm{D^{3}He}$ proton images. Figure \ref{fig:EP_Proton} shows
the comparison between the simulation synthetic and experimental $10.2\mathrm{MeV}$
TNSA proton image. Figure \ref{fig:prad_othercase} shows examples
of the proton images for $d=400\mathrm{\mu m}$ and $d=\mathrm{1200}\text{\ensuremath{\mu}m}$
case. The good qualitative agreement between the synthetic images
and the ones from experiment on the general trend of large scale features
suggests that the magnetic field structures we predict using FLASH
simulation are consistent with the structures in the experiments. 

Nernst effect can affect the evolution of magnetic field\citep{Nernst_Gao2015,Nernst_Lancia2014}.
This might be the reason why there are some disagreements in small
scale structure and sizes between the experimental data and the synthetic
images. In Table \ref{tab:Simulated-plasma-properties}, the product
of the electron gyro-frequency $\omega_{ce}$ and electron collision
time $\tau_{e}(=1/\nu_{e})$ is $\omega_{ce}\tau_{e}\sim15$ at $r=0$
and $\omega_{ce}\tau_{e}\sim9$ at $r=1\mathrm{mm}$. The value $\omega_{ce}\tau_{e}>1$
indicates that Nernst effect is important for our experiments. The
MHD model with Nernst term will be implemented in FLASH in the future.
We will make detail qualitative comparison in a future study with
Nernst term included.

\section{Conclusions and Discussions}

The FLASH simulation results were validated against a subset of experimental
data from the OMEGA experiments. The creation of the jets and strong
magnetic fields using the ring laser pattern is explained. 3D simulations
reproduce some features in previous 2D cylindrical results\citep{Fu_Fu2013,Fu_Fu2015}.
However, many new features emerges in 3D, e.g. the ``sun flower''
density pattern and the alternating para-axial magnetic field bundles.
Some questions still remain, e.g. the under-prediction in the line
width of IAW spectrum in Figure \ref{fig:ts_spect_compare}. An accurate
modeling for the magnetic fields requires implementation of Nernst
effect in FLASH code. Simulations using higher resolutions are also
desired. The XRFC modeling will be discussed in a future study.

The geometry of magnetic fields in our jets may be different from
the generally believed models in many astrophysical context, e.g.
the magnetic field of the jet along the axis of an accreting black
hole\citep{BH_Beckwith2008}, where the toroidal field supposedly
dominates. However, much can still be learned about the magnetic effect
on jet collimation, stability and structure in the laboratory. The
characteristics of the magnetized jet can be well controlled by tuning
the ring radius and increasing number of beams. By varying the hollow
ring radius, laser and target properties, we can achieve a large dynamic
range for the jet parameters, thus creating a highly versatile laboratory
platform for laser-based astrophysics. By using the jets we created,
shocks and shear flows can be studied with jet-jet collisions. 

The hollow ring laser platform is also ideally suited to scale up
to NIF with 192 beams and more energy per beam, creating centimeter-sized
magnetized jets. The jets produced with the NIF platform will have
several distinctive properties from OMEGA experiments, but are of
key importance for astrophysical jet modeling. The higher temperature,
density, flow velocity, and magnetic field will lead to large dimensionless
parameters. A turbulence regime is possible. The longer pulse on NIF
can sustain the jet for longer time, so that the radiative cooling
for doped targets become significant and useful to make the aspect
ratio larger. The aspect ratio can become large enough($\gg10$) that
the stability study can become more relevant to astrophysics. The
physical parameters of the jet can be tuned in such ways that various
collisionless and collisional regimes of the plasma can be accessed.
The dimensionless parameters for astrophysical jets may be better
realized in the large scale jets of NIF. Magnetic field geometry may
be tuned by increasing the number of beams.

{} 

\section{Acknowledgment}

This research is supported by DOE grant DE-NA0002721. The research
and materials incorporated in this work were partially developed at
the National Laser Users\textquoteright{} Facility at the University
of Rochester\textquoteright s Laboratory for Laser Energetics(LLE),
with financial support from the U.S. Department of Energy(DOE) under
Cooperative Agreement DE\nobreakdash-NA0001944. This work used the
Extreme Science and Engineering Discovery Environment (XSEDE\citep{XSEDE_Towns2014}),
which is supported by National Science Foundation(NSF) grant number
ACI-1548562. This research used resources of the National Energy Research
Scientific Computing Center, a DOE Office of Science User Facility
supported by the Office of Science of the U.S. Department of Energy
under Contract No. DE-AC02-05CH11231. Additional simulations were
performed at the Argonne Leadership Computing Facility and with Los
Alamos National Laboratory institutional computing. YL and EL acknowledge
partial support by LANL-LDRD during the writing of this paper. We
also acknowledge the valuable discussions with Hui Li.

{}

\bibliographystyle{apsrev4-1}
\bibliography{bibtex}

\end{document}